\documentclass[fleqn,usenatbib]{mnras}
\usepackage[utf8]{inputenc}
\usepackage{graphicx}
\usepackage{xcolor,ulem}
\usepackage{amssymb}
\usepackage{amsmath}
\usepackage{float}
\usepackage{stfloats}

\def\s{\mathrm{s}} 

\def\cm{\mathrm{cm}} 

\def\g{\mathrm{g}} 

\def\keV{\mathrm{keV}} 
\def\erg{\mathrm{erg}} 

\def\max{\mathrm{max}} 
\def\head{\mathrm{h}} 

\def\rad{\mathrm{rad}} 
\newcommand{\ch}{\mathrm{ch}}

\newcommand{\de}{\mathrm{d}}

\renewcommand{\epsilon}{\varepsilon}

\newcommand{\bo}{\mathrm{bo}} 

\newcommand{\jet}{\mathrm{j}}
\newcommand{\choke}{\mathrm{ch}}
\newcommand{\engine}{\mathrm{e}}
\newcommand{\volbo}{\sqrt{V_*/V_\bo}}

\definecolor{ao(english)}{rgb}{0.0, 0.5, 0.0}
\definecolor{applegreen}{rgb}{0.55, 0.71, 0.0}
\definecolor{darkpastelgreen}{rgb}{0.01, 0.75, 0.24}

\title[Choked Jets driven outflows]{The velocity distribution of outflows driven by choked jets in stellar envelopes}
\author{Matteo Pais}

\author[Pais, Piran \& Nakar]{
Matteo Pais,$^{1}$\thanks{E-mail: Matteo.Pais@huji.ac.il}
Tsvi Piran,$^{1}$\thanks{E-mail: Tsvi.Piran@huji.ac.il}
and Ehud Nakar$^{2}$\thanks{E-mail: udini@tauex.tau.ac.il}
\\
$^{1}$Racah Institute for Physics, The Hebrew University, Jerusalem 91904, Israel\\
$^{2}$School of Physics and Astronomy, Tel Aviv University, Tel Aviv 69978, Israel}

\date{November 2022}

\usepackage{graphicx}
\usepackage{newtxtext,newtxmath}

\begin{document}

\maketitle

\begin{abstract}
Many stripped envelope supernovae (SNe) present a signature of high-velocity material responsible for broad absorption lines in the observed spectrum. 
These include SNe that are associated with long  gamma-ray bursts (LGRBs) and low-luminosity GRBs ({\it ll}GRBs), and SNe that are not associated with GRBs. 
Recently it was suggested that this high velocity material originates from a cocoon that is driven by a relativistic jet. 
In LGRBs this jet breaks out successfully from the stellar envelope, while in {\it ll}GRBs and SNe that are not associated with GRBs the jet is choked. 
Here we use numerical simulations to explore the velocity distribution of an outflow that is driven by a choked jet and its dependence on the jet and progenitor properties. 
We find that in all cases where the jet is not choked too deep within the star, the outflow carries a roughly constant amount of energy per logarithmic scale of proper velocity over a wide range of velocities, which depends mostly on the cocoon volume at the time of its breakout. 
This is a universal property of jets driven outflows, which does not exist in outflows of spherically symmetric explosions or when the jets  are choked very deep within the star.
We therefore conclude that jets that are choked (not too deep) provide a natural explanation to the fast material seen in the early spectra of  stripped envelope SNe that are not associated with LGRBs and that properties of this material could reveal information on the otherwise hidden jets. 
\end{abstract}

\begin{keywords}
stars: jets -- gamma-ray burst: general --  supernovae: general -- hydrodynamics
\end{keywords}


\section{Introduction}

Both long and short GRB jets have to cross a significant amount of matter (the stellar atmosphere for long GRBs and the merger's ejecta in short ones) before producing the observed $\gamma$-rays.  
This understanding has lead to great interest in jet propagation within surrounding matter and the question was explored both analytically \citep[e.g.,][]{Blandford_Rees1974, Begelman_Cioffi1989, Meszaros_Waxman2001, Matzner2003, Lazzati_Begelman2005, Bromberg2011} and numerically \citep[e.g.,][]{Marti+1995, Marti+1997, Aloy+2000, macfadyen_supernovae_2001, Reynolds+2001, Zhang+2004, Mizuta+2006, Morsony+2007, Wang+2008, Lazzati+2009, Mizuta+2009, Morsony+2010, Nagakura+2011, Lopez+2013, Ito+2015, Lopez+2016, Harrison2018}.
This, naturally, raises the possibility that some jets are ``choked" during their propagation and are unable to break out of the surrounding dense medium. 
The observed temporal distributions of both long \citep{Bromberg+2012} and short \citep{Moharan_Piran2017} GRBs suggest that this happens in both types of events and there are indications that this also happens in some Supernovae \citep{Piran2019}.

In cases that the jet does not emerge we may still observed the signature of the cocoon that forms. 
First, the breakout of the shock driven by the cocoon produces a bright flash. 
For example, a cocoon breakout is most likely the origin of low-luminosity GRBs ({\it ll}GRBs) \citep{Kulkarni1998, macfadyen_supernovae_2001, Tan+2001, campana_association_2006, Wang+2007, waxman_grb_2007, katz_fast_2010, Nakar_Sari2012,Nakar2015}. 
These type of GRBs are rarely observed, however, when their low luminosity is taken into account it was realized that they are more numerous than regular LGRBs \citep{Soderberg+2006}. 
Another signature arises from the fast cocoon material that engulfs the star once the hot cocoon material breaks out and spreads. 
Specifically, this material leads to very broad absorption lines that are visible as long as it is optically thick \citep{Piran2019}. 
Such lines have been observed in several SNe some accompanied by {\it ll}GRBs  \citep{Galama+1998, Iwamoto+1998,   Modjaz+2006, Mazzali+2008, Bufano+2012, Xu+2013, Ashall+2019,Izzo_et_al_2019} and others without \citep{Mazzali+2000, Mazzali+2002, Mazzali+2009}. 
Finally, the cooling emission of the cocoon will also generate a potentially detectable UV-optical transient on time scale of hours to days \citep{nakar_piran2017}.

The important signature that helps determining the origin of the broad absorption lines is the energy-velocity distribution of the fast moving material. 
Regular spherical explosion result in a very steep  distribution  with roughly $\de E(v)/\de \ln v \propto v^{-5}$ \citep[e.g.,][]{nakar_sari2010}. 
However, when a jet is involved in the explosion this distribution is expected to be much shallower with much more energy at high velocities.
Recently, \cite{eisenberg2022} have shown that when the jet is successful the cocoon generates a unique energy-velocity flat distribution with $\de E/\de \ln \Gamma\beta \propto {\rm const.}$ over a wide range of velocities from sub to mildly relativistic, where $\beta=v/c$ and $\Gamma$ is the corresponding Lorentz factor. 
They also found that, when the jet is choked, it leaves a unique signature of a flat energy-velocity distribution. 
However, in the case of choked jets the flat distribution covers a range of velocities that is narrower than that of outflows driven by successful jets. 
Motivated by these results we examine here in detail the energy-velocity distributions of different chocked jet, focusing on the relation between the properties of the choked jet and the final energy-velocity distribution of the outflow after it becomes homologous.

For our study we use a large set of 2D relativistic hydrodynamical simulations. 
We consider explosions that are driven by choked jets in which we vary the opening angle and the engine working time of the jet as well as the structure of the progenitor. 
We follow the simulations until the entire outflow becomes homologous and examine what is the relation between these properties and the outflow energy-velocity distribution.

The paper is structured as follows. 
In Section \ref{sec: methodology} we describe the numerical procedure adopted for the simulations. 
In Section \ref{subsec: sim_setup} we describe the code choice and the composite mesh structure adopted, while in Section \ref{subsec: ics} we report in detail the setup for the stellar and interstellar environment and the initial conditions for the relativistic jet. 
Numerical aspects are discussed in two appendixes: a resolution study is described in Appendix \ref{sec: appendix A} and in Appendix~\ref{sec: appendix B} we explore the different use of a numerical smoothing function for the stellar density profile. 
In Section \ref{sec: results} we explore the results of our set of simulations.
We summarize our findings and consider the implications to  observations in Section \ref{sec: conclusions}.


\section{Methodology}
\label{sec: methodology}

\subsection{Simulation Setup}
\label{subsec: sim_setup}

Our simulations are performed using  the open source  massively parallel multidimensional relativistic magneto-hydrodynamic code {\textsc{pluto}} (v4.3) \citep{Mignone2007}. 
The code uses a finite-volume, shock-capturing scheme designed to integrate a system of conservation laws where the flow quantities are discretized on a logically rectangular computational grid enclosed by a boundary. 
We use the special relativistic hydrodynamics module in 2D cylindrical coordinates. 
We perform our calculations using a parabolic reconstruction scheme combined with a third-order Runge-Kutta time stepping. 
We also force the code to reconstruct the 4-velocity vectors at each time step. 

The  2D  simulations enables us to reach  high resolution  with reasonable computational resources. 
3D simulations carried by \citet{Harrison2018} suggest a similar generic evolution of the jet for the same parameters.
The propagation of the outer cocoon and the jet head are similar both in 2D and 3D simulations and the differences for the velocity distribution of the system are negligible - especially for choked jets - as shown in \citet{eisenberg2022}.
The main difference arises in the morphology of the  jet head in 2D simulations that is affected by a plug at the head front. 
This plug  diverts some of the jet elements sideways to dissipated their energy in oblique shocks but it is irrelevant for the cocoon structure in  which we are interested here. 
Another difference is the stability of the boundary between the jet and the inner cocoon \citep{Gottlieb_Nakar_Bromberg_2021}. 
However, all properties that are important for the cocoon (like mixing at the head, mixing between inner and outer cocoon and the propagation of the outer cocoon are all similar for 2D and 3D.   

Those differences are not significant for our purposes.

We chose the equation of state of the fluid to be ideal and with a constant relativistic polytropic index of $4/3$. 
This equation of state is applicable for a relativistic gas (as in the jet) as well to a radiation dominated Newtonian gas, such as the shocked stellar envelope.

To study the long term evolution of the jet and the cocoon from the star, we use a large grid spanning for several orders of magnitude. 
This  allows us to track the evolution of the system for at least two minutes after the breakout.  
At that time the entire stellar envelope is shocked by the cocoon and it expands enough to become homologous. 
We use a grid of size $4736 \times 4636$ cells, with the radial cylindrical coordinate\footnote{Throughout the paper $r$ is used for the 2D cylindrical radius while $R$ stands for the 3D radius.} 
extending within the range $r = [0,350] \times 10^{10} \cm$ and the vertical coordinate extending within the range $ z= [0.1 , 360] \times 10^{10} \cm$. 

We use a combination of a  uniform and two  non-uniform mesh grids in $r-z$ coordinates with a decreasing resolution from the inner region of the simulation box to the outer boundaries. 
The grid mesh is uniform in the inner part to maintain a high resolution of the jet injection and the formation of the resulting high pressure cocoon.
The uniform mesh has  $1000 \times 900$ grid points extending in the ranges $r = [0, 1] \times 10^{10} \cm$ and $ z=[0.1, 1] \times 10^{10} \cm$ with a resolution along both coordinates of $\Delta (r,z)_\mathrm{unif.} = 10^{7} \cm $.
Next to the uniform mesh we placed a stretched mesh with $1278^2$ grid points extending along both coordinates within the range $(r,z) = [1,6] \times 10^{10} \cm$ with a stretching ratio of $\sim1.0018$ 
The number of grid points for this mesh is chosen such that its initial grid spacing is the same of to the adjacent uniform mesh $\Delta (r,z)_\mathrm{s, init} = \Delta(r,z)_\mathrm{unif.} = 10^7 \cm$ and its final grid spacing is $\Delta(r,z)_\mathrm{s, final} = 10^8 \cm $. 
We cover the remaining grid  at larger distances with a logarithmic spaced mesh with $2458^2$ grid points extending within the range $(r,z) = [6, 360] \times 10^{10}\cm$. 
The number of grid points is chosen such that the  grid spacing of the mesh at $(r,z) = 6 \times 10^{10} \cm$  coincides to the resolution of the stretched mesh, such that $\Delta(r,z)_\mathrm{log, init} = \Delta(r,z)_\mathrm{s, final} = 10^8 \cm $. 
In this way we ensure a smooth increase of the resolution without jumps for the entire simulation grid.
A detailed resolution study for these simulations is reported in Appendix \ref{sec: appendix A}.

We inject the jet along  the inner  lower $z$ boundary, denoted $z_0$ (see \ref{sec:jet}). 
Otherwise, we impose a reflective  boundary condition at this boundary as it approximates the equatorial plane of the system. 
We impose  axial-symmetric conditions for the inner vertical boundary. 
Both outer boundaries are set to outflow.

\subsection{Initial conditions}
\label{subsec: ics}

\subsubsection{The star}

We approximate the stellar density profile  as a continuous power law that mimics the sharp decline of density in radius near the stellar edge\footnote{This profile diverges at the origin but this region does not influence the jet propagation and it is not included in our computational domain.}
\begin{equation}
\label{eq: rho_profile}
 \rho(R) = \begin{cases}
 \rho_* \left( \dfrac{R_*}{R} - 1\right)^2 + \rho_0 ,& \mathrm{for} ~  R \leq  R_* \ , \\ 
 \rho_0, & \mathrm{for} ~ ~ R > R_* \ .
 \end{cases}
\end{equation}
Here we choose $\rho_* = 100 ~ \g ~ \cm^{-3}$ and $R_* = 3\times 10^{10} \cm$. 
The total integrated mass of the star is $M_* = (9 \pi / 5) ~ M_\odot$ (see \ref{sec:scale} for scaling of these parameters to other values.)

For this  density profile  the local slope, $\alpha \equiv \de \log \rho(R)/\de \log R = {-2}/{(1-R/R_*)} $. 
The slope, $\alpha$, reaches the critical value of 3 for $R = R_*/3$. Beyond this value a spherical blast wave  accelerates and eventually looses causality. 
We present our results for this specific density profile. 
However, in Sec.~\ref{sec: diff_profiles} we show that the results for different stellar density profiles (both inner and outer) are qualitatively similar.

Surrounding the star we have an external CSM density of $\rho_0 = 1.67 \times 10^{-21} ~ \g ~\cm^{-3}$.
This exact value is unimportant as it is added just to avoid a numerical vacuum. 
The interaction of the jet or the cocoon outflow with this CSM is insignificant. 
To avoid numerical artifacts arising from the sudden drop in density at the edge of the star  we smooth the density of the outer edge of the star with a power law: 
\begin{equation}
\label{eq: smooth}
    \rho_\mathrm{smooth} (R) = \rho_\mathrm{s} \left( \dfrac{R}{R_\mathrm{s}} + 1\right)^{-8} \ , 
\end{equation}
with $ \rho_\mathrm{s} = 0.05 ~ \g ~ \cm^{-3}$ and a gradient scale of $R_\mathrm{s} = 5 \times 10^8 \cm$. 
We verified that this arbitrary choice of the smoothing function does not affect our results (see Appendix ~\ref{sec: appendix B}).
In order to avoid any initial random motion we set a uniform and low ambient pressure of $P = 3.5 ~ \keV ~ \cm^{-3}$ within the simulation grid. 

\subsubsection{The jet}
\label{sec:jet}

We inject a collimated jet with a constant luminosity $L_\jet$, operating for $t_\mathrm{e}$ so that the total injected energy is $E_{0} = L_\jet \times t_\mathrm{e} =  10^{51} \erg$. 
A uniform jet is injected through a nozzle with a velocity in the $z$ direction with an initial bulk Lorentz factor $\Gamma_{0,\jet}$ a  density $\rho_\jet $ and a specific enthalpy $h_\jet \gg 1$. 
Being relativistically hot the jet spreads quickly to form an initial opening angle  $\theta_\jet \simeq 1/(1.4 \Gamma_{0,\jet})$ (see details on this injection method at \citealt{Mizuta2013} and \citealt{Harrison2018}).

The jet is numerically initialized by the injection of density, pressure, and momentum along the $z-$direction through a nozzle parallel to the $z-$axis with a radius $r_\jet$ at an initial height $z = z_0$. 
The head cross section is then  $\Sigma_\jet = \pi r_\jet^2$. 
For an initial opening angle $\theta_\jet > 0.1~\rad$, we set up $r_\jet = 10^8~ \cm$, allowing a sufficient mesh coverage over the nozzle and we set the initial injection height at $z_0 = 10^9~\cm$. 

We consider a constant jet luminosity $L_\jet$.
This determines the product $\rho_\jet h_\jet$ as:
\begin{equation}
    \label{eq: rho_j}
    \rho_\jet h_\jet = \dfrac{L_\jet}{\Sigma_\jet  \Gamma_{0,\jet}^2 c^3} \ .
\end{equation}
We choose $h_\jet = 100$.  
This choice of the enthalpy is arbitrary, as long as $h_\jet \gg 1$.
The jet's pressure is given by $  P_\jet = (h_\jet - 1) {\rho_\jet c^2}/{4}$.

We explored the parameter space running simulations for different initial values of $L_\jet $ at steps of $2.5 \times 10^{50}~\erg~\s^{-1}$ from $2.5 \times 10^{50}~\erg~\s^{-1}$ to $2 \times 10^{51}~\erg~\s^{-1}$ for a total of 9 different luminosities. 
For each value of the luminosity, we run simulations for a set of different opening angles $\theta_\jet = [0.05, 0.1, 0.2, 0.4, 0.6]~ \rad$. 
As we keep the total jet energy fixed these conditions translate to different engine working times $t_\mathrm{e} = 10^{51}\erg / L_\jet$. 
For each of the 9 values of the luminosity we run 5 different values of the opening angle for a total of 45 simulations.

\subsubsection{Scaling  relations}
\label{sec:scale}
While we consider specific numerical values for the stellar and jet parameters, our solutions can be scaled to other values. 
The equation of motion of the forward shock speed $\beta_\head$, is regulated by  $\Tilde{L}$ \citep{Matzner2003,Bromberg2011}:
\begin{equation}
\label{eq: tilde_L2}
    \Tilde{L} \simeq  \dfrac{L_\jet}{ \Sigma_\jet \rho(R) c^3} \ ,
\end{equation}  
with
\begin{equation}
\label{eq: beta_h}
    \beta_\head = \dfrac{1}{1+{\Tilde{L}}^{-1/2}} \ .  
\end{equation}
The stellar size $R_*$, is the  scale  length of t the system. 
The scalings $\Sigma_\jet \propto R_*^2$ and $\rho(R/R_*) \propto \rho_*$ we can express $\tilde{L}$ as
\begin{equation} 
\Tilde{L} \propto \dfrac{E_0}{t_\engine \rho_* R_*^2} \ . 
\end{equation}
If we scale the  stellar  radius as $R_* = \lambda R_*'$ we have to scale the density and the jet luminosity accordingly in order to maintain $\Tilde{L}$ and $\beta_\head$ unchanged. 

As we show later the location where the jet is choked (i.e. where the last element launched by the jet reaches the head) with respect to the stellar radius has also to be constant. 
The choking location $z_\choke$ is roughly proportional to the engine time $t_\engine$ \citep{Nakar2015}:
\begin{equation}
\label{eq: zchoke}
    z_\choke = \int_0^{t_\choke} \beta_\head c \de t \simeq \beta_\head c t_\choke = \dfrac{\beta_\head c }{1-\beta_\head} t_\engine  \ .  
\end{equation}
where $  t_\choke =  {t_\engine}/{(1 - \beta_\head)}$ is the choking time. 
If $\beta_\head$ is  kept constant than any transformation on $R_*$ will leave $z_\choke / R_*$ unchanged if $t_\engine = \lambda t_\engine' \propto R_*$. 
Thus, scaling $E_0 = \eta E_0'$ and $R_* = \lambda R_*'$ we require that and $\rho_* = \eta \lambda^{-3} \rho_*'$.

Because $M_* \propto \rho_* R_*^3$, when keeping this scaling of $t_\engine$ and $R_*$, we can rewrite Eq.~\ref{eq: tilde_L2} as $ \Tilde{L} \propto E_0/M_*$, which means that since $\Tilde{L}$ is kept  constant the typical  velocity of the system $v_0 = (2E_0/M_*)^{1/2}$ is also conserved under these transformations.

Turning to the jet parameters we recall that the only relevant quantities are $L_\jet$, $t_\engine$ and $\theta_\jet$.
The first two determine $E_0$ and the latter determined $\Gamma_{0,\jet}$. 
The luminosity, together with the stellar parameters determine the produce $\rho_\jet h_\jet$ with the condition that $h_\jet$ while arbitrary should be much larger than one. 

In summary, given the physical scales $R_*$, $\rho_*$, $E_0$, $v_0$, $t_\engine$ and the scalings $R_* = \lambda R_*'$, $E_0 = \eta E_0'$, the parameters defining the physics of our system, i.e. $\tilde{L}$, $z_\ch/R_*$, will not change if  $t_\engine = \lambda t'_\engine$ and $\rho_* = \eta \lambda^{-3} \rho'_*$.

\section{Results}
\label{sec: results}
\begin{figure*}
    \centering
    \includegraphics[scale=0.352]{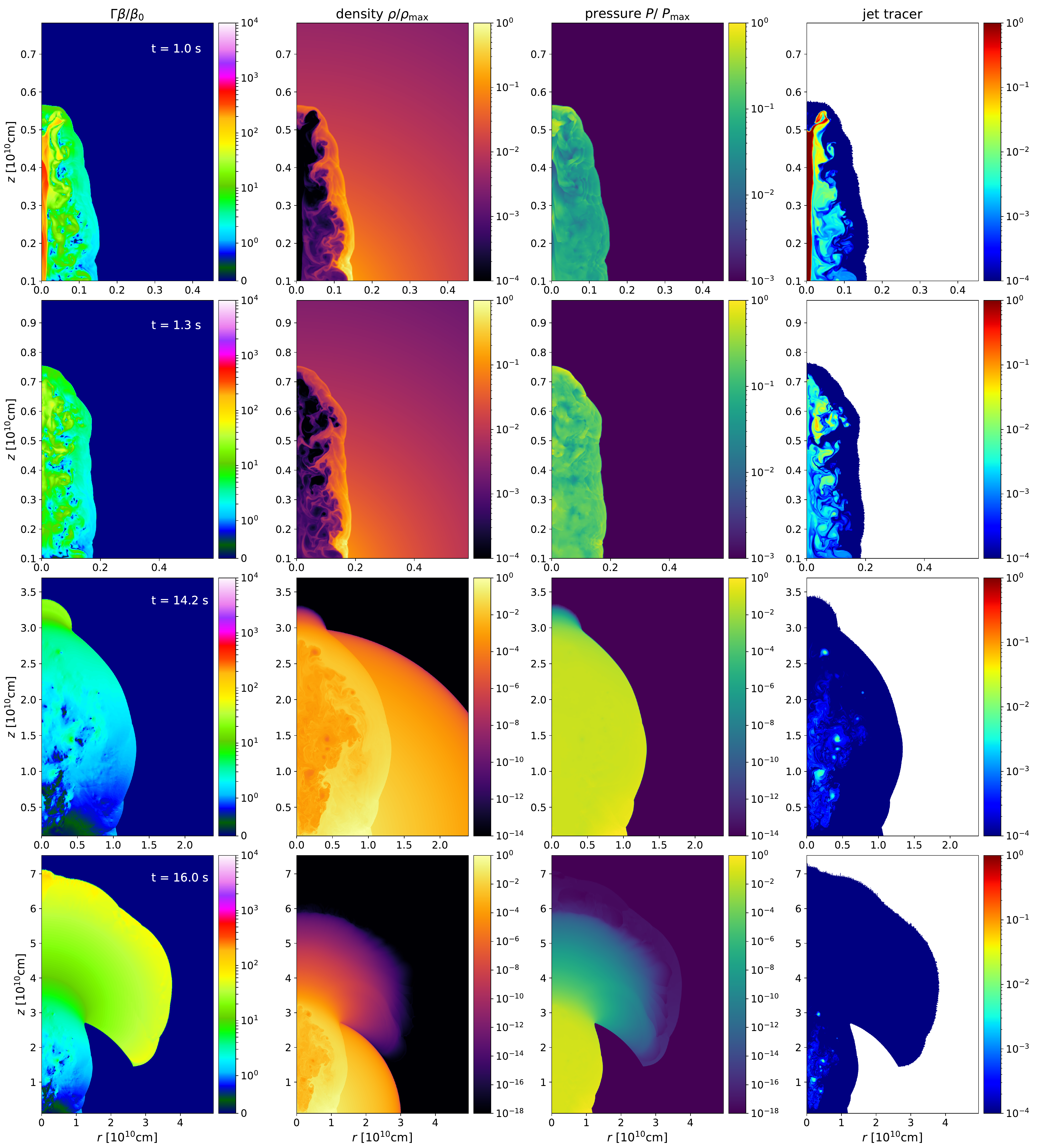}
    \caption{A canonical jet ($\theta = 0.2$ rad; $L_\jet = 10^{51} \erg \ \s^{-1}$ and $t_e=1 \s$)  launched in a test star with a density profile given by Eq.~\ref{eq: rho_profile}. 
    The four panels show, from left to right, the relativistic $\Gamma\beta$ factor, the density $\rho$, the pressure $P$ and a scalar tracer of the ejected jet material (an unitary value implies pure jet material with no mixing). 
    All the quantities but the $\Gamma\beta$ factor are normalized to the respective maximum values in order to increase the color contrast. The scale for the velocity of our system is dictated by $v_0$. With the above parameters  $\beta_0 =v_0/c =  0.014$ and the relativistic regime begins when $\beta \Gamma/\beta_0\approx 50$.
    The different rows show the evolution of the jet at $t = t_\engine = 1~\s$ (when the engine stops, first row) - the jet is clearly seen here surrounded by a cocoon, $t=1.3~\s$ (choking time, second row) - the jet has disappeared as its tail reached the head, $t=14.2~\s$ (shortly after the breakout, third row), and $t=16.0~\s$ (sideways spreading outside the star, fourth row). 
    To enhance the color contrast we change the normalization scale of $\rho/\rho_\max$ and $P/P_\max$ in the third and fourth row in order to capture the tenuous material and pressure spilling outside the star. }
    \label{fig: jet_t01}
\end{figure*}

\begin{figure*}
    \centering
    \includegraphics[scale=0.35]{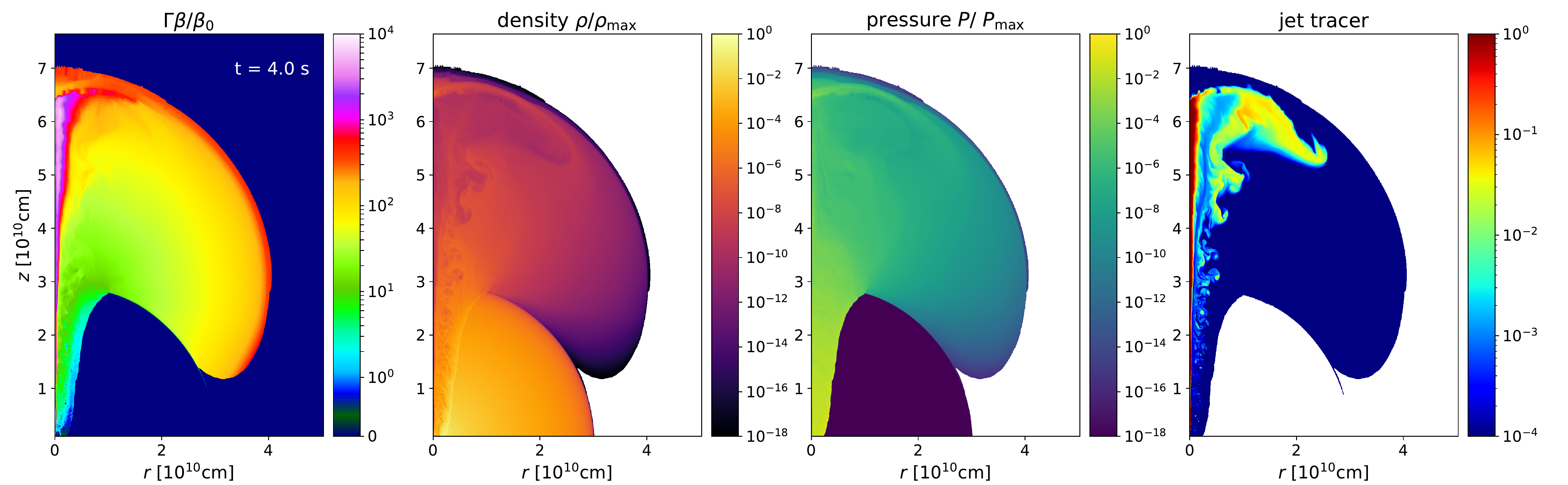}
    \caption{4-panel figure of the breakout of an unchoked jet at $t=t_\engine = $4 s with an opening angle of $\theta_\jet = 0.1$ rad with a luminosity of $L_\jet=2.5 \times 10^{50}~\erg~\s^{-1}$. 
    In this case the jet tail did not catch up the jet head, resulting in an unchoked breakout and with a head propagation operating as the jet engine were still active. We use the same color scale of the fourth row of Fig.~\ref{fig: jet_t01} for comparison.}
    \label{fig: jet_t4}
\end{figure*}

\begin{figure*}
    \centering
    \includegraphics[scale=0.38]{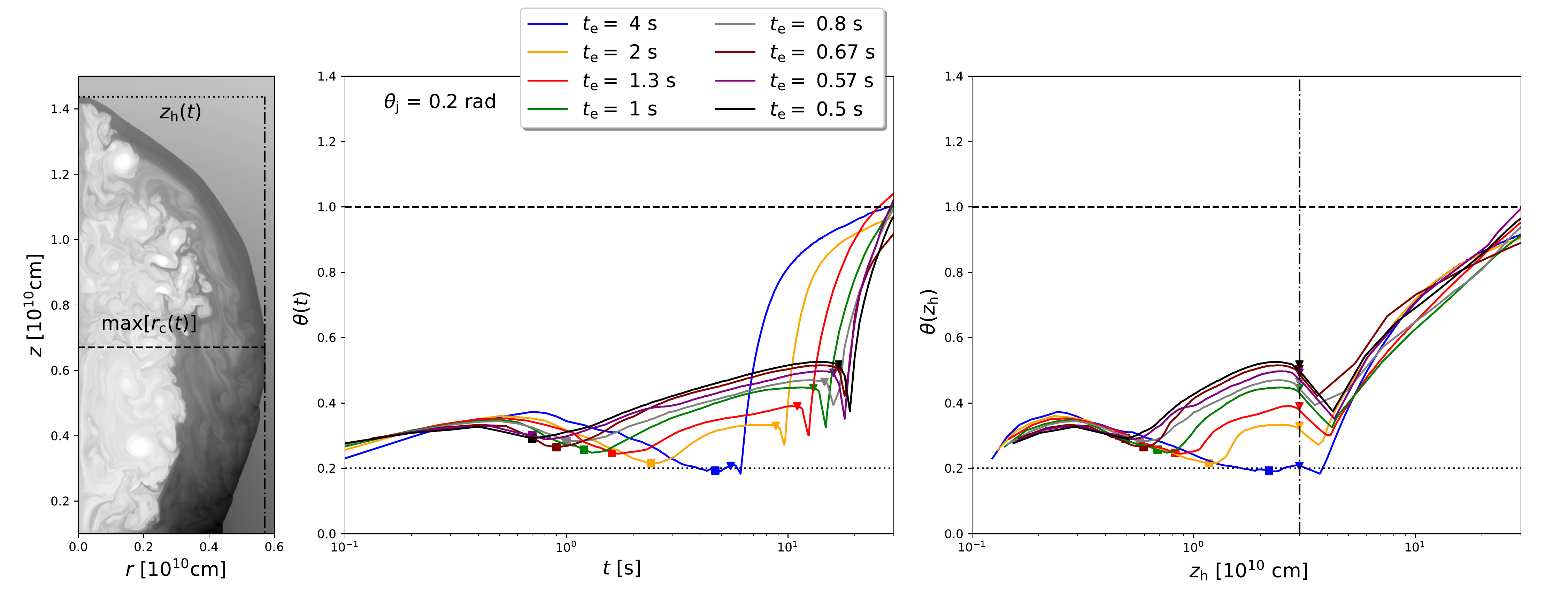}
    \caption{\emph{Left}: A schematic figure of how the aspect ratio $\theta(t)$ is defined (Eq.~\ref{eq: theta_of_t}) superposed on the density map of the jet cocoon while it is still inside the star. 
    The dashed line represents the maximal width of the cocoon, while the dashed-dotted line represents the maximal height. 
    \emph{Centre and right}: Evolution of the aspect ratio of the jet as a function of time (central panel) and of the cocoon/jet head location $z_\head$ (right panel). 
    The square dots represent the choking time of the jet while the triangular dots represent the breakout time of the cocoon. 
    In both frames the thick dashed horizontal line represents the condition for an isotropic blast wave ($r_\mathrm{c} = z_\head$), while the horizontal dotted line represent the original opening angle of the launched jet ($0.2~\rad$ for the figure). 
    On the right-hand side the dashed-dotted vertical line represents the star edge, located at $R_* = 3 \times 10^{10}~\cm$.}
    \label{fig: spreading_angle}
\end{figure*}

\begin{figure*}
    \centering
    \includegraphics[scale=0.35]{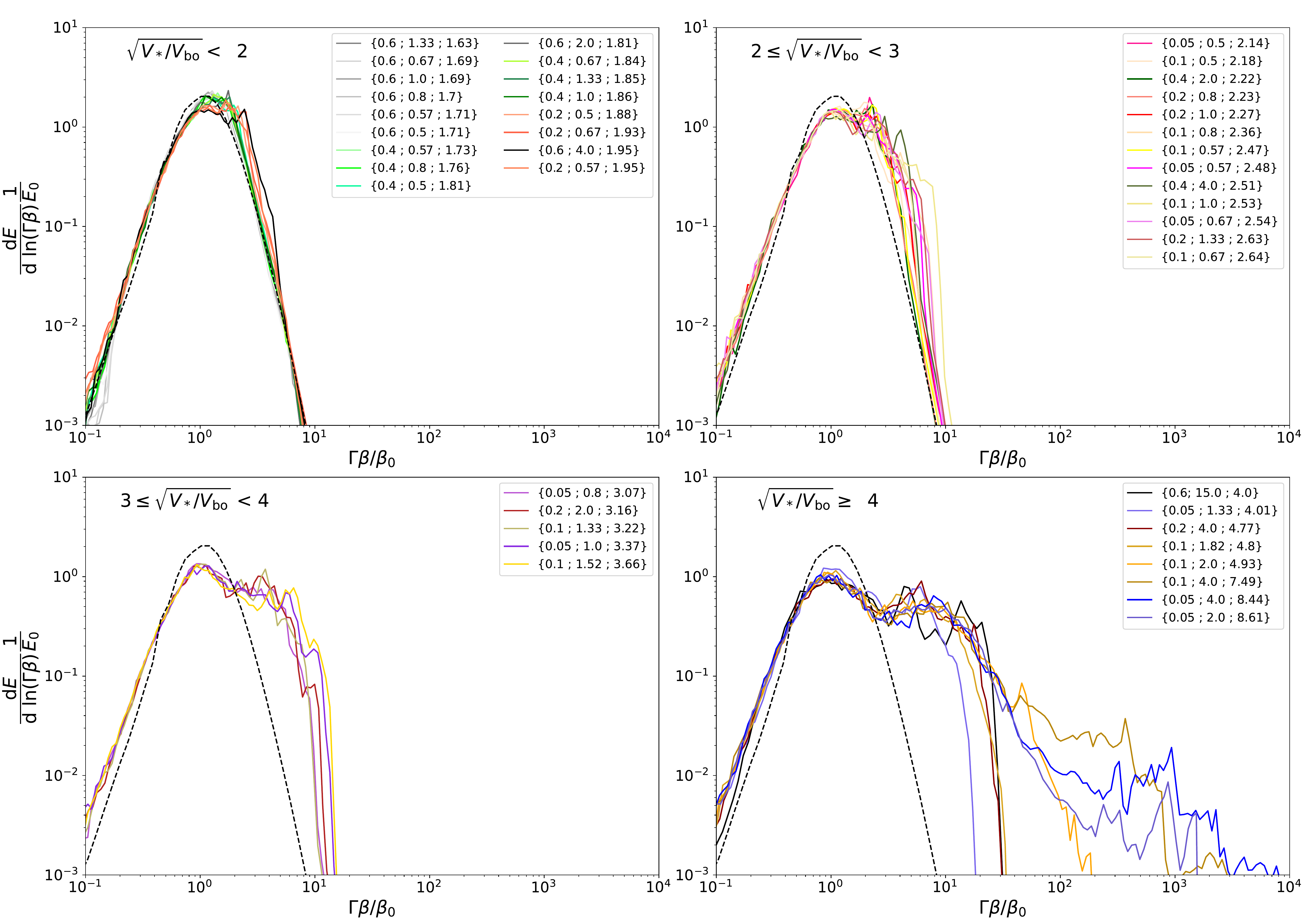}
    \caption{Classification of the energy-velocity distribution grouped according to their different cocoon volume $V_\bo$ at $t=t_\bo$. 
    Each curve is marked by a different shade of color and a triplet of numbers indicating, in order: the opening angle in radians, $t_\engine$ in seconds, and then $\sqrt{V_*/V_\bo}$. The dashed black line represents the isotropic case.}
    \label{fig: choking_height}
\end{figure*}

\begin{figure}
    \centering
    \includegraphics[scale=0.34]{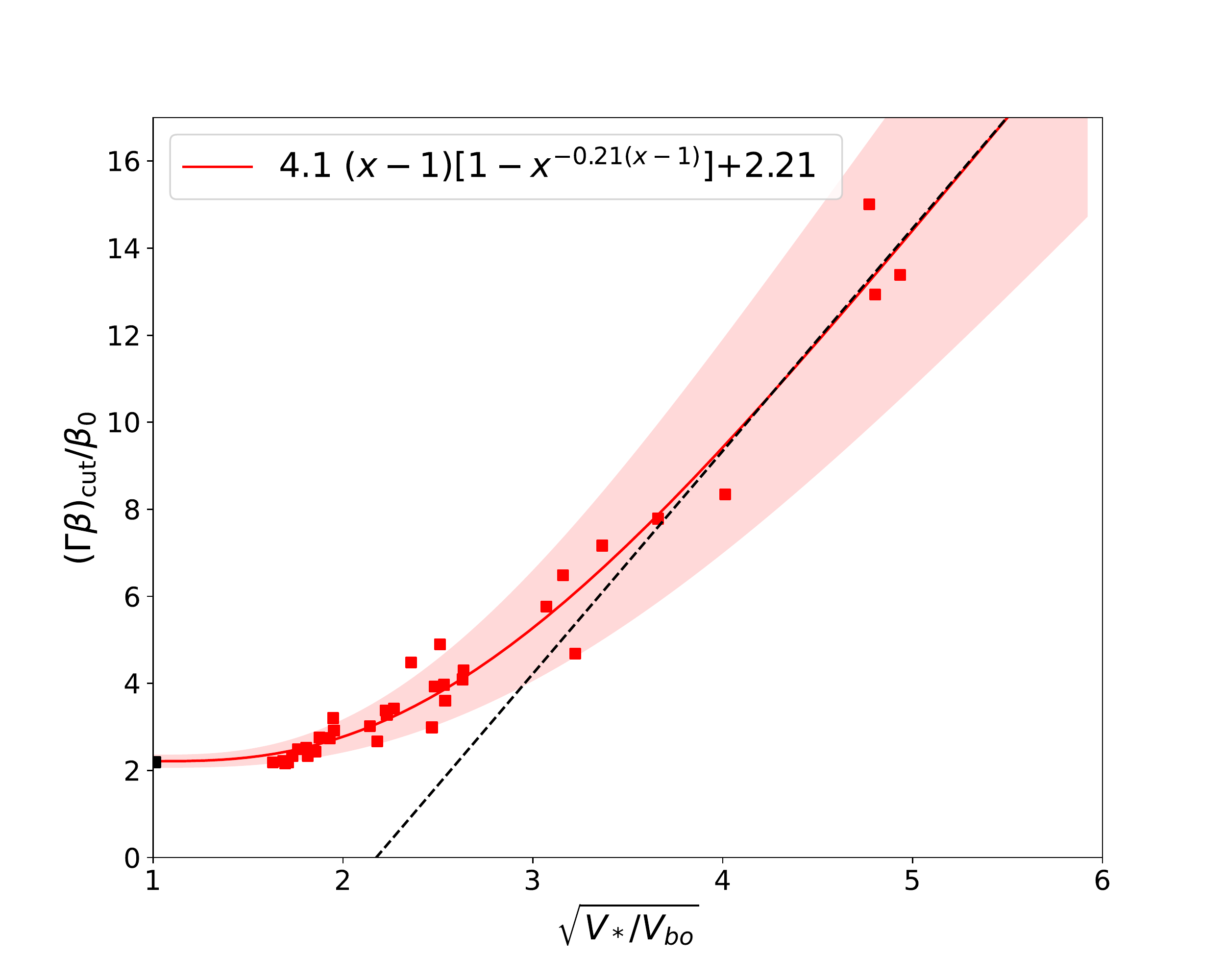}
    \caption{Correlation between $\sqrt{V_*/V_\bo}$ and the cutoff of the energy-velocity distribution for the simulated set of jets for a  cutoff value of  $0.25$ of the maximum of each differential energy distribution in the plots of Fig.~\ref{fig: choking_height}. 
    The red-colored area represents 1-standard deviation error of the red fit curve. 
    From the fitting formula we see that the distribution cutoff corresponds to 4 times the the square root of the volume ratio at the breakout for values of $(\Gamma\beta)_\mathrm{cut}$ above 3. 
    The black dashed line represents the linear limit for high cutoffs.}
    \label{fig: volume_cutoff}
\end{figure}

\begin{figure}
    \centering
    \includegraphics[scale=0.34]{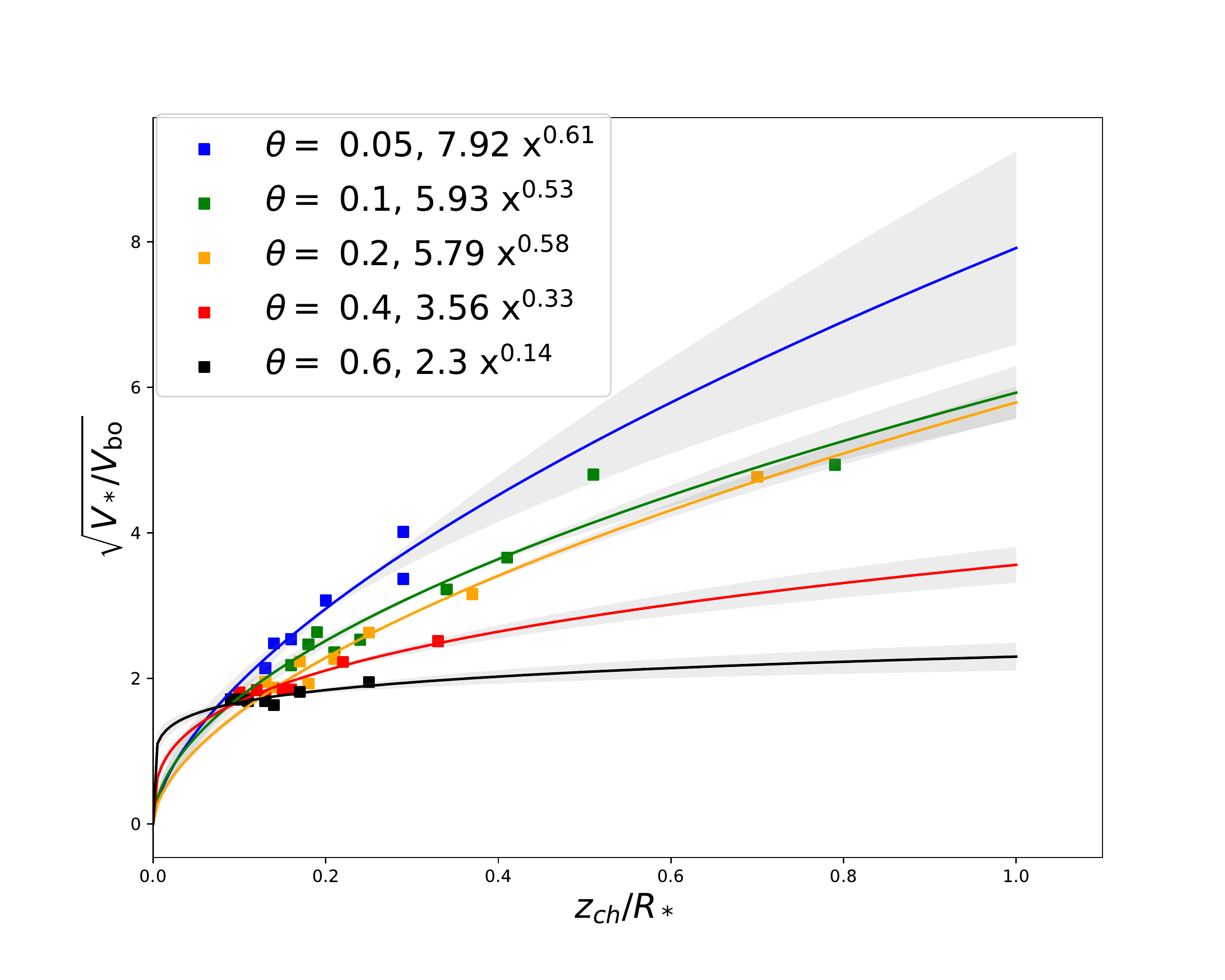}
    \caption{Correlation between the cocoon volume at the breakout and the choking height of the jet for different values of the initial opening angle $\theta_\jet$. 
    The gray-shaded region represents the 1$\sigma$ error for the fits. }
    \label{fig: v_bo_z_ch}
\end{figure}

\subsection{The jet-cocoon system}
\label{subsec: analysis}

We start analyzing our simulation set considering a jet with our \emph{canonical} parameters of 1-sided luminosity of $L_\jet = 10^{51} \erg/\s$ and $\theta_\jet = 0.2~\rad \simeq 10^\circ$.

While advancing through the stellar atmosphere the interaction of the relativistic jet with the stellar material results in a forward-reverse shock structure that is called the head of the jet \citep{Blandford_Rees1974, Begelman_Cioffi1989, Meszaros_Waxman2001, Matzner2003, Lazzati_Begelman2005, Bromberg2011}. 
The jet head propagation  velocity, $\beta_\head$, is much lower than the jet velocity before it reaches the head and for typical GRB jets it is Newtonian. 
The shock-heated jet and stellar material that enters the head flow sideways because of the high head pressure and form a pressurized  cocoon which enshrines the jet.  
The contact discontinuity between the material shocked in forward and the reverse shocks divides the cocoon to  inner and outer parts. 
The inner cocoon is composed of tenuous jet material which has crossed the reverse shock while the outer cocoon is composed of denser shocked stellar material. 
The cocoon exerts a pressure on the jet such that, if sufficiently high,  collimates it, thus reducing its opening angle and consequently reducing the jet cross section compared to the uncollimated jet. 

Within our chosen stellar structure model the jet head moves at a constant velocity at the inner region of the core where the local density slope $\alpha = 2$.  
If the jet reaches outer regions where $\alpha > 2$  it starts accelerating.

The jet is \emph{choked} if the engine stops while the jet is propagating within the stellar envelope and the last jet element launched by the engine (\emph{tail}, hereafter) catches up with the jet head before the latter breaks out of the star.
In this case all the engine's energy goes into the cocoon. 
Clearly, the choking height  (Eq.~\ref{eq: zchoke}) satisfies $z_\choke < R_*$. 
Otherwise we define the jet as unchoked or successful. 
Throughout the following analysis we focus mostly on jets choked at various depths inside the star. 
For comparison we show also the case of an unchoked jet breaking out of the star.
For a detailed study on the energy-velocity distribution of stellar explosions which are driven by successful jets see \cite{eisenberg2022}.

We divide the evolution of the jet to three different phases: 1) the injection phase and choking phase $t < t_\choke$, 2) the cocoon expansion phase $t_\choke < t < t_\bo$ and 3) the cocoon breakout phase $t > t_\bo$.  
The different phases of a choked jet are shown in Fig.~\ref{fig: jet_t01}.

\subsubsection{Injection and Choking: $t \leq t_\choke$}
The engine operates for $t_\engine$ producing a jet. 
This is clearly seen in the first row of Fig.~\ref{fig: jet_t01} in both the rightmost panel showing $\beta \Gamma$ and the leftmost one showing that the tracer of the jet material is concentrated mostly within a narrow cylinder along the symmetry axis $z$ with radius $r \simeq r_\jet$ (color dark red).
This behaviour is typical of the collimated regime \citep{Bromberg2011}.

After the jet engine stops the last jet material launched at the injection nozzle propagates upwards. 
At $t_\engine$ the jet head is still unaware that the engine stopped and the jet head continues to propagate with $\beta_\head$ (in this specific simulation $\beta_\head \simeq 0.2)$ after the central jet engine is switched off.
However, as $\beta_\head < 1$ while the jet material moves at $\beta_\mathrm{t} \simeq 1$ the jet tail catches up with the head.
Only at that time the information that the engine stopped reaches the head and the reverse shock within the jet disappears. 
This is the time where the jet is choked.  

As the head propagates with a velocity of $\beta_\mathrm{h} c$ and the tail propagates at $\beta_\mathrm{t} \simeq 1$, we can estimate the the time that the jet tail will catch up with the head  at  $t_\choke$, as defined in Eq.~\ref{eq: zchoke}.
Until $t < t_\choke$ the jet continues to drive the head forward through the stellar atmosphere. 
The second row of Fig.~\ref{fig: jet_t01} shows the  system  at $t = t_\choke$, roughly 0.3 seconds after the end of the engine activity, in this specific simulation. 
One can see that the very fast jet material around the core disappeared. At this stage all the jet's energy has been dissipated and given to the surrounding cocoon. 

\subsubsection{Cocoon expansion: $ t_\choke < t < t_\bo$}

After the jet choking, the cocoon becomes less and less collimated  and proceeds spreading sideways while the forward shock decelerates when it is deep within the envelope and accelerates as it reaches the steep density gradient near the stellar edge \citep{Irwin_Nakar_Piran2019}.
During the propagation the inner cocoon transfers energy to the freshly shocked material (via PdV work).

\subsubsection{Brekout: $t> t_\bo$}
After the breakout the cocoon material spreads both radially and tangentially to engulfs the stellar surface, quickly shrouding the breakout point from most observers (see  the last row of Fig.~\ref{fig: jet_t01}). 
The star is blanketed by the ejecta in a time equal to $t_\mathrm{wrap} \simeq \pi R_* /2 v_\mathrm{bo}$, where $v_\mathrm{bo}$ is the breakout velocity of the cocoon near the pole.
The shock driven by the cocoon also moves tangentially towards the equator at a slower pace until the entire stellar envelope is shocked at $t_\mathrm{shock} \simeq \pi R_* /(2 v_\mathrm{p})$ where $v_\mathrm{p}$ is the pattern velocity at which the spilled material travels along the stellar surface \citep{Irwin_et_al2021}. 
Shortly after reaching  the equator the shocked material propagates almost radially and outwards and it becomes homologous once the outflow reaches $\sim 2R_*$. 

\subsubsection{Successful jets} 
Jets whose engine operates long enough break out from the stellar envelope before the end of the activity of the central engine. 
These jets are not choked and can preserve an ultra-relativistic velocity once they get out of the star. 
We show an example of a successful jet in Fig.~\ref{fig: jet_t4}. 
Because the jet broke out without being choked the cocoon structure inside the star is mostly collimated along the vertical axis. 
From the first and fourth panel which show $\Gamma\beta$ and the jet tracer respectively, it is evident how the innermost region is still dominated  by tenuous, highly relativistic jet material. 
Comparing the last row of Fig.~\ref{fig: jet_t01} and Fig.~\ref{fig: jet_t4} we notice that for the same normalized density and normalized pressure scale, the longer duration of the unchoked jet results in expulsion of denser and faster stellar material with respect to the choked jet. 

\subsection{The spreading angle and the cocoon volume}

To describe quantitatively the geometry of the jet-cocoon during its propagation within the stellar envelope we use the aspect ratio, defined as
\begin{equation}
\label{eq: theta_of_t}
    \theta(t) = \dfrac{\max(r_\mathrm{c}(t))}{z_\head(t)}\ ,
\end{equation}
where $r_\mathrm{c}$ is the cocoon cylindrical radius and $z_\mathrm{h}$ is the head position. 
For $\theta \ll 1$ the aspect ratio is a good approximation of the cocoon spreading angle.
The expanded cocoon at the moment of the breakout is shown in the third row of  Fig.~\ref{fig: jet_t01}. 
The steep density transition results in an elongation and acceleration of the cocoon and the ejection of low-density material from the star, which rapidly engulfs the star's external layers. 
We define the breakout angle $\theta_\bo$ as the geometric opening angle measured at the breakout time $t=t_\bo$, namely 
\begin{equation}
\label{eq: theta_bo}
    \theta_\bo = \theta (t_\bo) = \dfrac{\max(r_\mathrm{c}(t_\bo))}{R_*}\ .
\end{equation}

The evolution of $\theta(t)$ for $\theta_\jet = 0.2~\rad$ and several values of $t_\engine$ is reported in Fig.~\ref{fig: spreading_angle}. 
At first, immediately after injection, the aspect ratio starts growing. 
The growth continues until $z_\head$ is roughly twice the injection radius, $z_0$, at which point the aspect ratio starts decreasing, approaching the point where the cocoon opening angle is comparable to $\theta_\jet$. 
This evolution reflects the time it takes the pressure in the cocoon to build up to the point that it starts collimating the jet effectively (see \citealt{Harrison2018} for details). 
The evolution of the aspect ratio changes dramatically immediately as the jet is fully choked. 
Since there is no more fresh jet material to drive the head its velocity drops sharply. 
At the same time the cocoon pressure, and thus its sideways expansion, is not affected. 
The result is that the aspect ratio grows continuously after $t_{\ch}$. 
There is a short episode, just before and after the breakout when the aspect ratio decreases, as the head accelerates near the edge of the star and after the breakout. 
Soon after that the aspect ratio starts increasing rapidly as some of the material that broke out of the star spreads sideways at speed that is close to the speed of light. 
One clear property that is seen in the figure is that jet that are choked more deeply have longer time to expand before they breakout and therefore a deeper choking results in a wider cocoon with a larger volume at the time of breakout. 
As we show next this fact has important implications for the energy-velocity distribution of the outflow.  

The volume of the cocoon at breakout, $V_\bo$  is another parameter that describes the  properties of the jet-cocoon system. 
As the energy of a choked jet is given to the cocoon, for a given energy, the cocoon mass (and hence volume) at breakout, will corresponds to a typical expansion velocity of the cocoon material. 
As the volume-averaged density in the shocked cocoon material and the volume-averaged density of the star are roughly the same, we can define a characteristic velocity at the breakout linked with the breakout volume, namely:
\begin{equation}
\label{eq: volume_bo}
    \beta_\bo \simeq \beta_0 \sqrt{\dfrac{V_*}{V_\bo}} \ . 
\end{equation}

\subsection{The Energy-velocity distribution}

Fig.~\ref{fig: choking_height} depicts the energy-velocity distribution of the entire set of simulations for different values of the engine working time $t_\engine$ and different initial opening angles $\theta_\jet$ at $t=120 \s$ (when the outflow is homologous and kinetic energy dominates). 
The $x$-axis is normalized by $\beta_0$ and the $y$-axis by $E_0$. 
Each curve is differentiated by color and labeled by a triplet of numbers describing, respectively, $\theta_\jet$, $t_\engine$, and $\sqrt{V_*/V_\bo}$.  
We  grouped the different curves according to $\sqrt{V_*/V_\bo}$. 
For a comparison we superposed the energy-velocity distribution of an isotropic spherically symmetric explosion (black-dashed line) for all panels. 

Fig.~\ref{fig: choking_height} shows, first, that in all cases the energy-velocity distribution exhibits a roughly constant energy per logarithmic scale of $\Gamma\beta$ over a range of velocities. 
The distribution rises quickly before this rough plateau starts and decays sharply after it ends. 
The rough plateau always starts at $\beta_0$ and its highest velocity is determined almost entirely by $V_{\bo}$, with a weak dependence on the jet opening angle. 
To estimate the highest velocity of the flat part of the distribution we define $\beta_{\rm cut}$ as the velocity obtained when the energy-velocity distribution drops to 1/4 of its maximum value. 
This arbitrary definition provides a velocity that is slightly larger than the end of the plateau (e.g., in the spherical case $\beta_{\rm cut}=2\beta_0$).
Fig~\ref{fig: volume_cutoff} shows that there is a strong positive correlation between $\beta_{\rm cut}$  and $\sqrt{V_*/V_{\bo}}$. 
For small values of $\sqrt{V_*/V_{\bo}}<3$, where typically $z_{\choke} \ll R_*$, we see that $\beta_{\rm cut}\approx \beta_{\bo}$. 
However, for larger values of $\sqrt{V_*/V_{\bo}}$ where the choking takes place not very deep within the stellar envelope, $\beta_{\rm cut} > \beta_{\bo}$. 
The origin of the material faster than $\beta_{\bo}$ in these cases is the inner cocoon, which retain a significant fraction of its energy at the time of the breakout and outer cocoon material that is close to the edge of the star, where the forward shock is faster than $\beta_{\bo}$.

The value of $V_*/V_{\bo}$ is expected to depend  on the jet opening angle and the choking depth. The jet opening angle determines the aspect ratio of the cocoon as long as the head that is pushed ahead by the jet is feeding the cocoon ($\theta\approx \theta_\jet$), while the choking depth determines by how much this aspect ratio increases until the breakout.  
Fig.~\ref{fig: v_bo_z_ch}  depicts the correlation between $V_\bo$ and $z_\choke$ for different values of the initial $\theta_\jet$. 
As expected, $V_\bo$ is a function of $z_\choke$ and $\theta_\jet$. 
A deeper choking height and a wider jet correspond to a larger  cocoon volume upon breakout.

\begin{figure}
    \centering
    \includegraphics[scale=0.64]{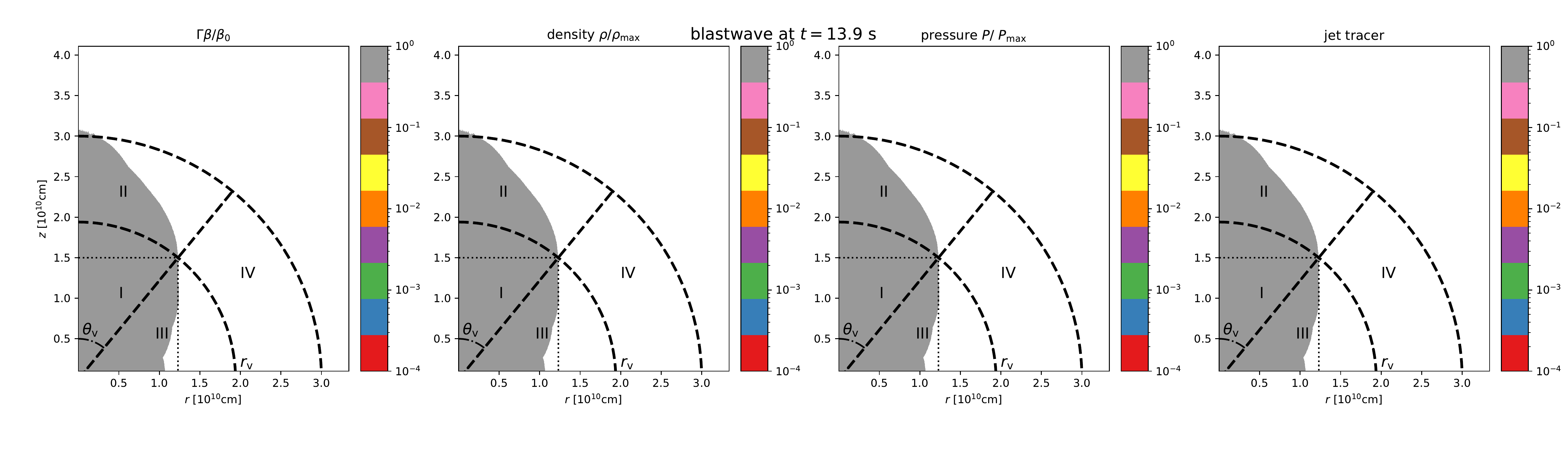}
    \caption{A sketch of the division of the stellar volume for the analysis of the distribution of the stellar material. 
    The cocoon shape taken at $t=t_\bo$ is overlaid. We associate a scalar tracer to each of the four sectors: I) internal-axis, II) external-axis, III) internal-equatorial, and IV) external-equatorial. }
    \label{fig: scheme}
\end{figure}

\begin{figure*}
    \centering
    \includegraphics[scale=0.34]{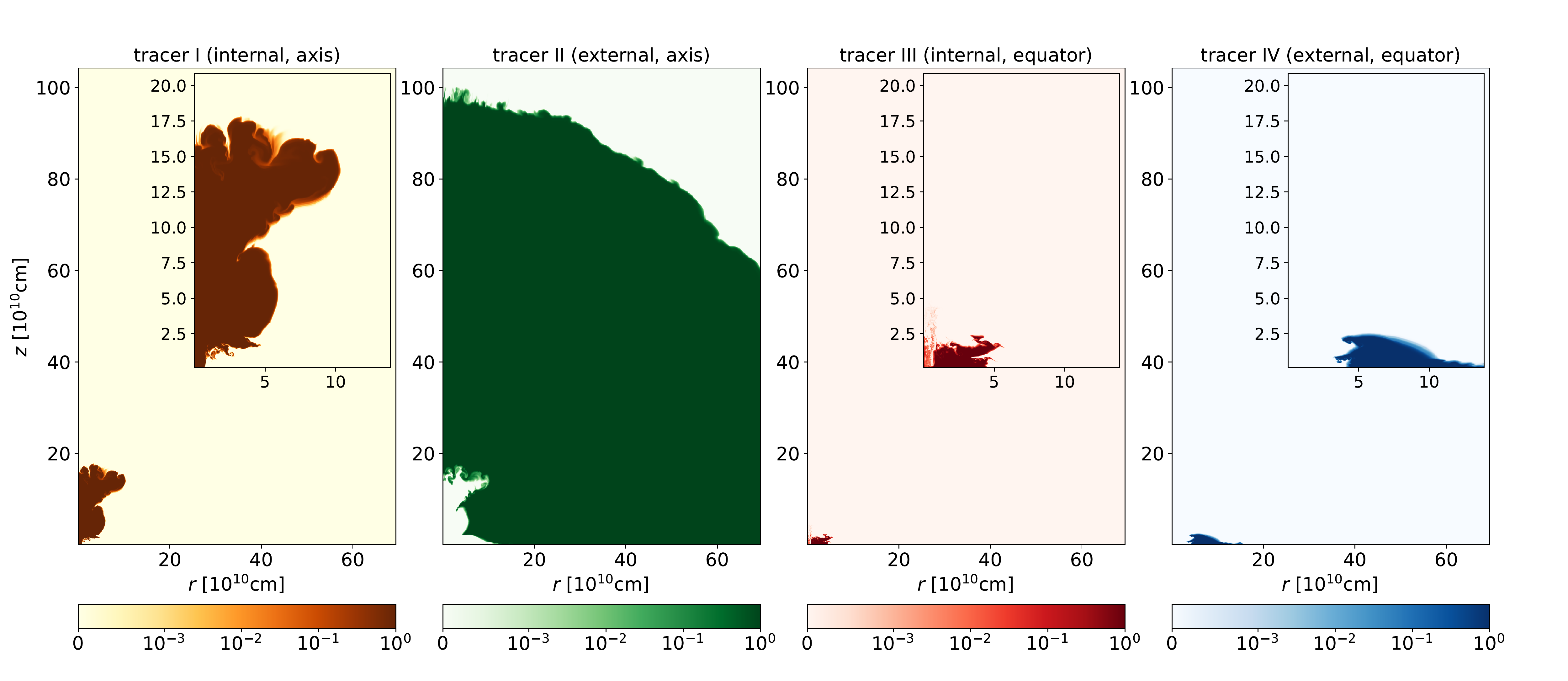}
    \caption{Maps of the four tracers I, II, III and IV (from left to right) associated with the four sectors of the stellar material (see Fig.~ \ref{fig: scheme}). 
    The maps show the distribution of the stellar material at $t=60~\s$ resulting  from  a jet with canonical parameters (see Sec~\ref{subsec: analysis}). }
    \label{fig: 4_tracers}
\end{figure*}

\subsubsection{The origin of ejecta with different final velocities}

\begin{figure}
    \centering
    \includegraphics[scale=0.51]{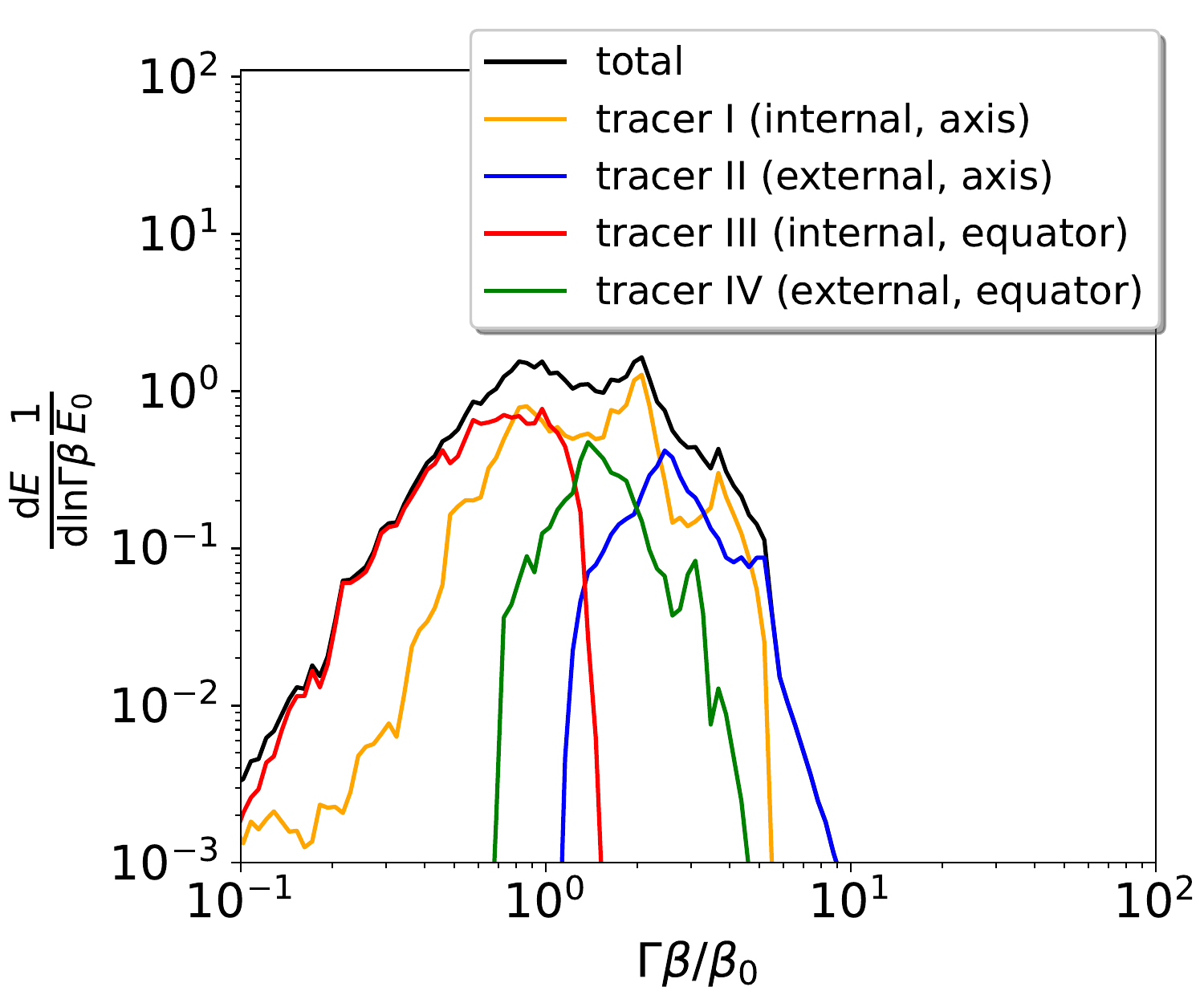}
    \caption{The energy-velocity distribution of the four different matter tracers associated with the sectors depicted in Fig.~\ref{fig: 4_tracers}.
    Matter from region II (external along the axis) dominates the highest velocity region. 
    Matter from region III (internal equatorial) dominates the low velocity regime. 
    Matter from region I (internal along the axis) dominates the intermediate region and the plateau. 
    Matter from IV (external equatorial) is always subdominant.}
    \label{fig: e_vs_v_tracers}
\end{figure}

To understand the origin of the various components of the outflow, we tracked the distribution of the ejected material using four different scalar tracers associated with four distinct regions of the star. 
The division to the different regions is determined at the time of the breakout and it is shown in Fig.~\ref{fig: scheme}. 
The tracers follow the mass in each of this region (at the time of breakout): I) internal-axis, II) external-axis, III) internal-equatorial, IV) external-equatorial.

Fig.~\ref{fig: 4_tracers} shows the distribution of the stellar material from each of the regions at $t=60~\s$, roughly $46~\s$ after the breakout.
Fig.~\ref{fig: e_vs_v_tracers} shows the energy-velocity distributions of the four sectors.

We see that the  quasi-spherical outflow that leads the ejecta is made only of tenuous material coming from the on-axis, external layers of the progenitor directly above the expanding jet cocoon (region II). 
This component contains only around $2\%$ of the total stellar mass but it contains 11\% of the total ejecta energy. 
Evidently, the fastest ejecta is dominated by this sector. 
The  material associated with the stellar core part that is along the axis (region I; first panel in Fig.~\ref{fig: 4_tracers}) is much more concentrated than that of the external-axis region (II) but much more extended than the two equatorial sectors. 
It contains $30\%$ of the total stellar mass and 46\% of the outflow energy. 
This section dominates the energy distribution over a wide range of velocities. 
Almost all the rest of the mass and the energy are contained in the internal-equatorial sector (III) which carries $60\%$ of the mass and 32\% of the energy. 
It dominates the energy at low velocities $\lesssim \beta_0$. 
Finally, the outer-equatorial section carries  $5\%$ of the ejecta mass and 11\% of its energy. 
All its material is moving at intermediate velocities and it is subdominant at all velocities. 

\subsection{The effect of the stellar density profile}
\label{sec: diff_profiles}

\begin{figure*}
    \centering
    \includegraphics[scale=0.43]{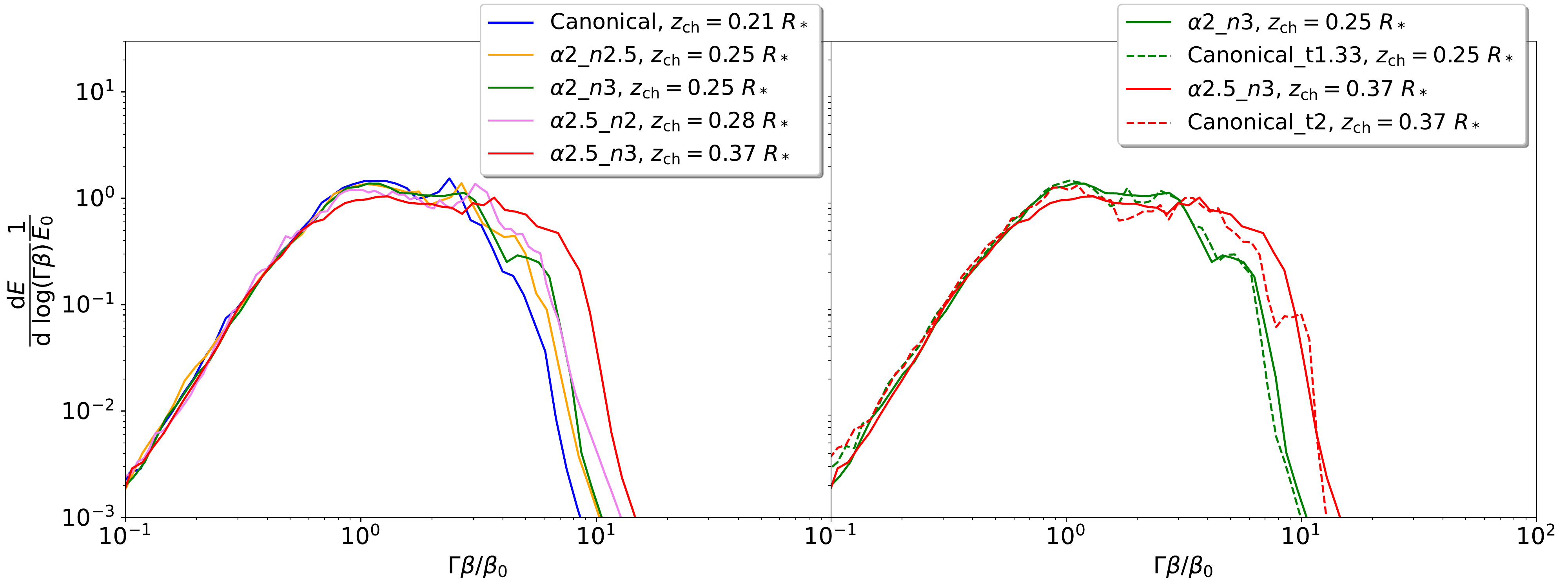}
    \caption{\emph{Left}: Energy-velocity distributions of  jet simulations with canonical parameters for four different stellar density profiles. 
    The blue line ($\rho \propto R^{-2} (R_*-R)^2 $) represents the profile used in most of the previous simulations. \emph{Right}: A comparison of the  energy distributions of the cases $\alpha=2,n=3$ and $\alpha=2.5,n=3$ with those arising from from two jets choked in a star with a canonical density profile at the same heights $z_\choke/R_*$, respectively.}
    \label{fig: density_profiles_s}
\end{figure*}

\begin{center}
\begin{table*}
\label{tab: different_profiles}
\begin{tabular}{|c||ccccc|}
\hline 
Jets & $t_\engine$ [s] & $\theta_\jet$ [rad] & $\rho(r)$ & $z_\choke/R_*$ & $t_\bo$ [s] \\ 
\hline 
Canonical & 1 & 0.2 & $\propto R^{-2} (R_*-R)^2 $ & 0.21 & 13.9 \\ 
$\alpha2$\_$n2.5$ & 1 & 0.2 & $\propto R^{-2} (R_*-R)^{2.5} $ & 0.25 & 11.2 \\ 
$\alpha2$\_$n3$ & 1 & 0.2 & $\propto R^{-2} (R_*-R)^3 $ & 0.25 & 9.5 \\ 
$\alpha2.5$\_$n2$ & 1 & 0.2 & $\propto R^{-2.5} (R_*-R)^2 $ & 0.28 & 6.8 \\ 
$\alpha2.5$\_$n3$ & 1 & 0.2 & $\propto R^{-2.5} (R_*-R)^3 $ & 0.37 & 4.0 \\ 
\hline 
Canonical\_t1.33 & 1.33 & 0.2 & $\propto R^{-2} (R_*-R)^2 $ & 0.25 & 11.5 \\ 
Canonical\_t2 & 2 & 0.2 & $\propto R^{-2} (R_*-R)^2 $ & 0.37 & 8.3 \\ 
\hline
\end{tabular}
\caption{Properties of the jets injected in different density profiles. The table lists the engine working time $t_\engine$, the initial opening angle $\theta_\jet$, the density profile $\rho(r)$ used in the run, the choking height relative to the star radius $z_\choke/R_*$, and the breakout time $t_\bo$.}
\end{table*}
\end{center}

To study the effect of different stellar density profiles we consider stellar density profiles that can be written as:
\begin{equation}
\label{eq: rhogen}
    \rho (R) = \rho_*\left(\dfrac{R_*}{R}\right)^\alpha  \left(1-\dfrac{R}{R_*}\right)^n \ ,
\end{equation}
where $n$ is the outer slope at the edge, and $\alpha$ is the inner slope, with $\alpha <3$. 
The density profile described by Eq.~\ref{eq: rho_profile}, which is used through the rest of the paper, is roughly equivalent to the case of $\alpha=2,~ n=2$ and it will be referred as the \emph{canonical profile} hereafter. 
The profiles that we consider are listed in Table.~\ref{tab: different_profiles}. For each profile we run a simulation with our canonical jet parameters, $\theta_\jet = 0.2~\rad$, $L_\jet = 10^{51}~\erg~\s^{-1}$, and we inject the jets from the same initial height ($z_0 = 10^9~\cm$). 

Fig.~\ref{fig: density_profiles_s} shows a comparison of the energy-velocity distributions from different stellar profiles. 
First, it shows that the distributions are all flat over a range of velocities, implying that them main feature of the outflow from an explosion driven by a choked jet is independent of exact stellar profile (a similar result was found by \citealt{eisenberg2022} in the case of explosions that are driven by successful jets). 
When looking in more detail, the right-hand side shows two pairs of simulation. 
Each pair shows the results of different stellar profiles with similar $\theta_\jet$ and $z_\choke$ (which dominates $V_{\bo}$). 
The distributions found in the two simulations of each pair are very similar, implying that when the cocoon properties are similar the stellar profile has a minor effect on the outflow energy-velocity distribution. 

On the left-hand side of Fig.~\ref{fig: density_profiles_s} we compare the energy-velocity distributions of jets with the exact same parameters (including $t_\engine$) but different envelope density profiles. 
It shows that the stellar profile affects the velocity of the head (as was found previously by \citealt{Bromberg2011,Harrison2018}) and therefore jets with the same properties are choked at different heights when propagating in different density profiles. 
Since the energy-velocity profile depends strongly on $z_\choke$ two jets with the same properties that propagate at different stellar profiles will result in outflows with different energy-velocity distributions, as shown on the right-hand panel of this figure.   

\subsection{The energy velocity distribution at different viewing angles}
\label{sec: profiles_different_angles}

\begin{figure*}
    \centering
    \includegraphics[scale=0.39]{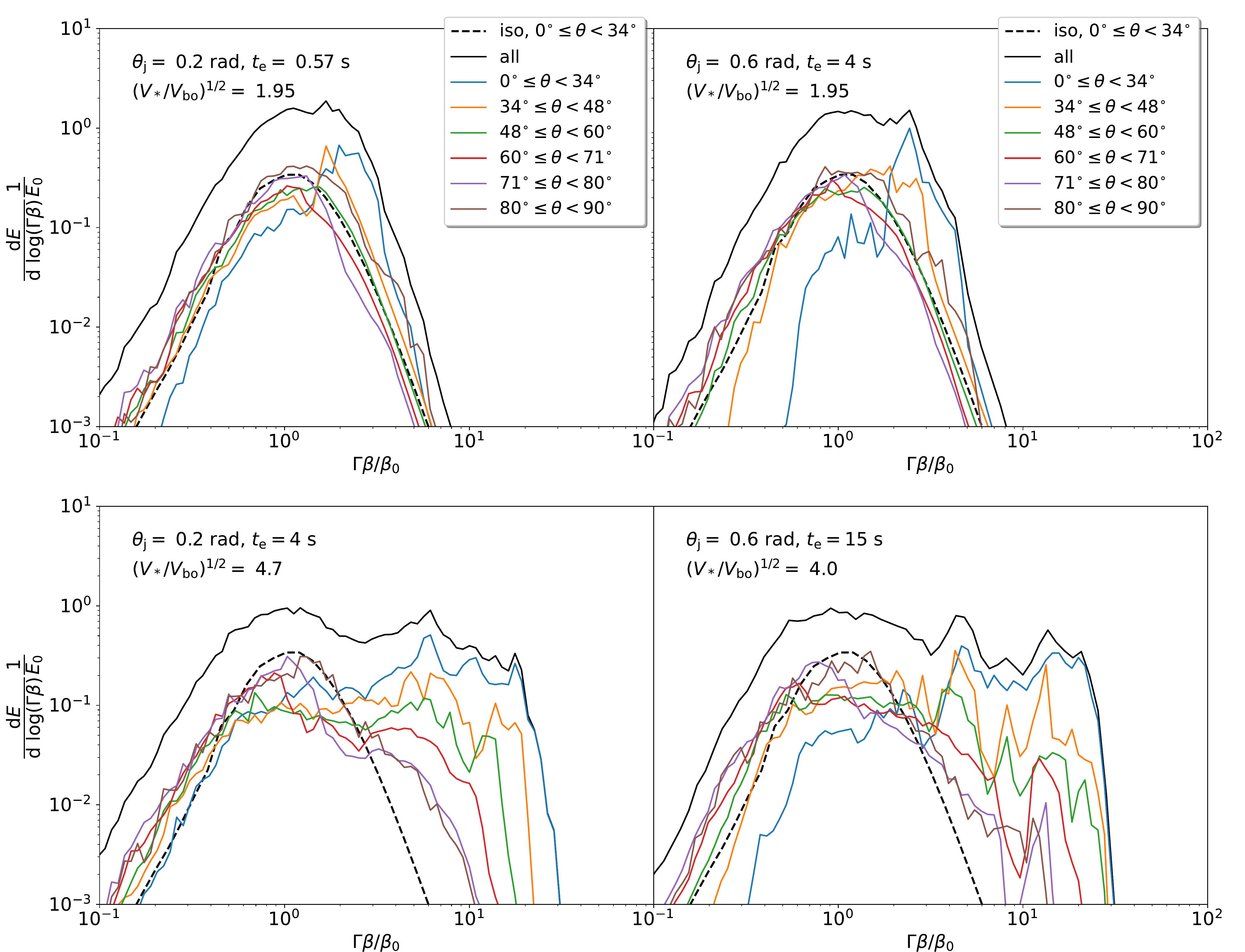}
    \caption{Energy-velocity distributions for six different viewing angles (chosen so that the corresponding wedges have the same volumes). \emph{Top}: Two jets with the same choking height $ (V_* / V_\bo)^{1/2}= 1.95 $  but different initial opening angles. 
    \emph{Bottom}: Two jets choked at $ (V_* / V_\bo)^{1/2} \sim 4 $ (bottom panels) and with the same opening angles of the jets on the top row, respectively. 
    The black dashed line represents the isotropic energy-velocity distribution for $1/6$ of the total volume. The profiles are taken at $t=120~\s$.}
    \label{fig: energy_profiles_volumes}
\end{figure*}

Since jet driven explosions are aspherical, one expects that the outflow will not be isotropic. 
Fig.~\ref{fig: energy_profiles_volumes} depicts the energy-velocity distribution of four simulations. 
For each simulation we show the distributions at six different sections, where each section is the sum of the ejecta within a range of polar direction. 
To see the dependence on the initial conditions we show simulations with two different jet opening angles ($0.2$ and $0.6$ rad) and two different values of breakout volume $\volbo$. 
As expected, the outflow is aspherical. 
A common, also expected, property of all four simulations is that the maximal velocity of the outflow is around the jet axis at lower polar angles. 
This result was found also for jet-driven explosions of successful jets \citep{eisenberg2022}. 
In the two simulations with the small value of $\volbo$ (i.e., low $z_\choke$ and/or wide $\theta_\jet$; top panels) the energy-velocity distribution of the equatorial outflow ($\theta \gtrsim 60^\circ$) is similar to that of a spherical explosion, with a typical velocity $\beta_0$. The faster outflow is confined to lower angles. 
In the two simulations with the large value of $\volbo$ (i.e., high $z_\choke$ and narrow $\theta_\jet$; bottom panels) a large range of velocities was seen in all directions, but still faster velocities are observed closer to the jet axis.

\begin{figure*}
    \centering
    \includegraphics[scale=0.42]{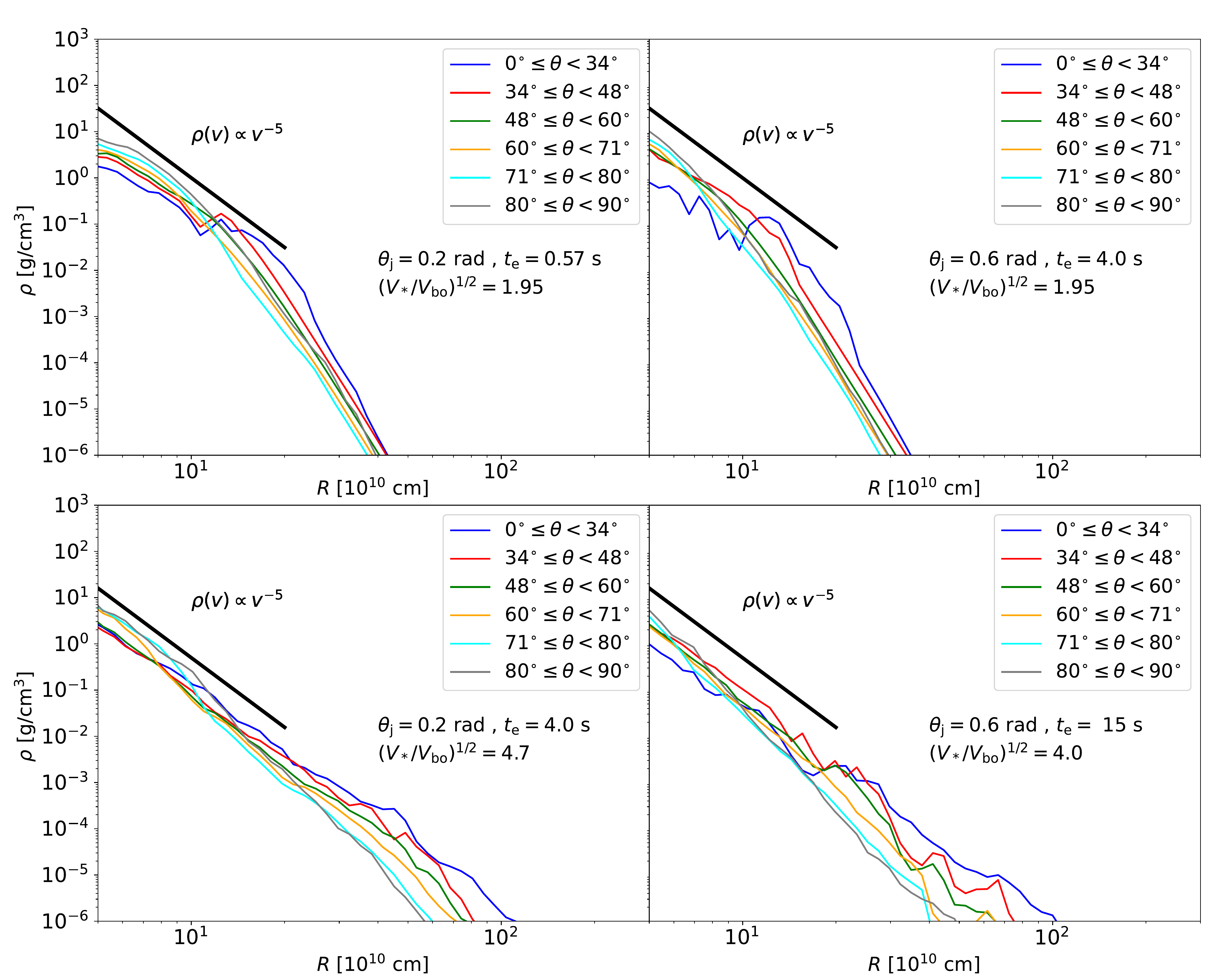}
    \caption{Density profiles at different viewing angles of two jets with a similar breakout volume $ (V_* / V_\bo)^{1/2}= 1.95 $ (top row) but different initial opening angle compared to the density profile of jets choked at $ (V_* / V_\bo)^{1/2} \sim 4 $ (bottom row) with the same opening angles, respectively. 
    The solid thick black line in each panel represent a power-law fit of $\rho(r) \propto v^{-5}$, corresponding to a flat distribution of energy per logarithmic bin of the velocity. 
    The profiles are taken at $t=120~\s$.}
    \label{fig: density_profiles_1}
\end{figure*}

For clarity we present in Fig.~\ref{fig: density_profiles_1} the radial density distribution profiles, $\rho(R)$, for different viewing angles of four different simulations. 
These are the same simulations and the same divisions to angular sections as in Fig.~\ref{fig: energy_profiles_volumes}. 
This presentation is often used in studies of SN ejecta and at sub-relativistic velocity  $\rho(R) \propto \beta^{-5} \frac{\de E}{\de \log\beta}$.


\section{Conclusions and implications to observations}
\label{sec: conclusions}

We carried relativistic hydrodynamical simulations in 2D cylindrical coordinates of stellar explosions driven by jets, focusing on configurations of choked relativistic jets and exploring  how those can lead to different realizations of the velocity distribution of the outflow in its homologous expansion phase.  
We followed the evolution of a relativistic jet from the injection deep inside the star to the point where it is choked and then continued to follow the cocoon as it emerges from the envelope and ultimately unbinds it up to the point that the outflow becomes homologous. 
We scrutinized the various stages of the jet inside the star and analyzed what happens during the choking process and the adiabatic cocoon expansion.  
While the results are given for a specific set of parameters we provided scaling relation for the physical parameters of the jet and the star in order to facilitate a dimensionless treatment of the problem. 
We stress that the scaling laws presented in Sec.~\ref{sec:scale} don't involve gravity and they are  valid as long as the gravitational binding energy of the star is subdominant with respect to the total energy of the jet. A conditions that holds for the powerful jets that have been observed in some SNe \citep{Piran2019}.

We summarize our findings as follows:

\begin{itemize}
    \item All jet driven explosions in which the jet is not choked too deep within the star generate an outflow with a unique feature: a significant range of velocities over which the outflow carries a roughly constant amount of energy per logarithmic scale of the proper velocity ($\Gamma\beta$). 
    This is a universal property of jet driven explosions. 
    The main difference between different setups is the range of velocities over which the energy is constant.
    
    \item The plateau of the energy-velocity distribution starts in all cases at $v_0=\sqrt{E_0/M_*}$. 
    The maximal velocity of the plateau depends mostly on the cocoon volume upon breakout and the corresponding velocity is  $\beta_\bo=\beta_0\sqrt{V_*/V_\bo}$. 
    For  $\sqrt{V_*/V_\bo} < 3$ the maximal velocity is comparable to $\beta_\bo$, while for larger values of $\sqrt{V_*/V_\bo}$ the maximal velocity is larger than $\beta_\bo$ and it can become mildly relativistic. 
    
    \item The volume of the cocoon upon breakout, $V_\bo$, depends on the choking height, $z_\choke$, and on opening angle of the jet upon launching. 
    A higher $z_\choke$ and narrower opening angle leads to a smaller $V_\bo$ and thus to an outflow that extends to higher velocities.
    
    \item The outflow from an explosion driven by a choked jet is not isotropic.
    In general, the material along the poles (that is along the jet direction) is faster while the material along the equator is slower.
\end{itemize}

A spherical explosions accelerate only a negligible fraction of the stellar mass to very high velocities. 
Indeed,  hydrodynamic simulations of core-collapse supernova explosions have proved to be rather aspherical without necessarily harboring a jet \citep[e.g.,][for a review]{Burrows_core_collapse_2006, Janka_2007, Janka_review_2012, Janka_review_2016}. However,  the differential mass-velocity distribution in those kind of simulations \citep{Wongwathanarat_2015} does not show a significant fraction of the ejected material with sufficient velocity to mimic the high-velocity tail seen in some SNe and produced by the jets considered here.

We have shown here that the situation is drastically different when there is a jet that breaks the symmetry. 
Such a jet can deposit a significant amount of energy at high velocity matter, even in case that the jet is choked within the envelope. 
This excess in high velocity outflow (compared to a spherical explosion) is certainly expected when the entire stellar explosion is driven by a jet, but it is also expected if the jet is accompanied by a simultaneous more spherical explosion  (see e.g., \citealt{eisenberg2022}).
If sufficiently optically thick such a high velocity material that surrounds a SN would produce a very broad  absorption lines (with typical width corresponding to 0.1-0.2c) in the observed spectrum. 
It will be observed in the early spectra but will disappear later when this outer envelope that is rapidly expanding becomes optically thin. 

Lines that show an excess of high velocity material have been observed in several SNe \citep{Galama+1998, Iwamoto+1998, Mazzali+2000,  Mazzali+2002, Modjaz+2006, Mazzali+2008, Bufano+2012, Xu+2013, Ashall+2019,Izzo_et_al_2019}.
Our result show that, as suggested by \cite{Piran2019},  a chocked jet can lead to that high velocity material. 
However, we have found that some conditions are needed to observe the corresponding broad absorption lines. 
First, the jet must be chocked at sufficiently large distance at the stellar atmosphere. 
The signature of jets that are chocked too deep will not be so significant.
Second, as there is less fast moving material in directions far from the jet direction, the fast moving matter will become optically thin earlier in these directions. 
As the broad absorption line will fade faster, this implies that observers at such viewing angles are less likely to observe the broad absorption line signature. 
These last two facts imply that we may not observer broad emission line in all SNe that harbour relativistic jets. 

The excess in fast material was observed in various types of stripped envelope SNe. 
This include SNe that are associated with long GRBs, SNE that are associated with {\it ll}GRBs and SNe that are not associated with GRBs at all. 
Long GRBs must contain successful relativistic jets. 
{\it ll}GRBs contain jets which may very well be choked \citep{Kulkarni1998, macfadyen_supernovae_2001, Tan+2001, campana_association_2006, Wang+2007, waxman_grb_2007, katz_fast_2010, Nakar_Sari2012,Nakar2015}. 
We do not know if SNe that are not associated with GRBs harbour jets, but if they are then these jets must be choked ones. 
A previous study by \cite{eisenberg2022} have shown that successful jets can generate the energy-velocity distribution which is observed in SNe that are associated with long GRBs. 
Our finding here show that choked jets can explain the energy-velocity distribution seen in SNe that are associated with {\it ll}GRBs and in SNe that are not associated with any type of GRBs. 
This provides further support for the interpretation of the ``disappearing'' early very broad absorption lines in some SNe as arising from choked jets. 
These findings also show that such lines may not be detected in all SNe that harbor choked jets. 
Further exploration of this model, including estimates of the observed spectra and the fraction of events in which these lines will be observed will be carried out in future work.

\section*{Acknowledgments}
We kindly thank Christopher Irwin for the stimulating discussions and suggestions. 
We also thank our anonymous referee for a constructive and helpful report.
This work is supported by the ERC grants TReX (TP and MP) and JetNS and an ISF grant 1995/21 (EN).
\section*{Data Availability}
The data underlying this article will be shared on reasonable
request to the corresponding author.

\bibliographystyle{mnras}
\bibliography{main}

\appendix 
\section{Resolution Check}
\label{sec: appendix A}

\begin{figure}
    \centering
    \includegraphics[scale=0.33]{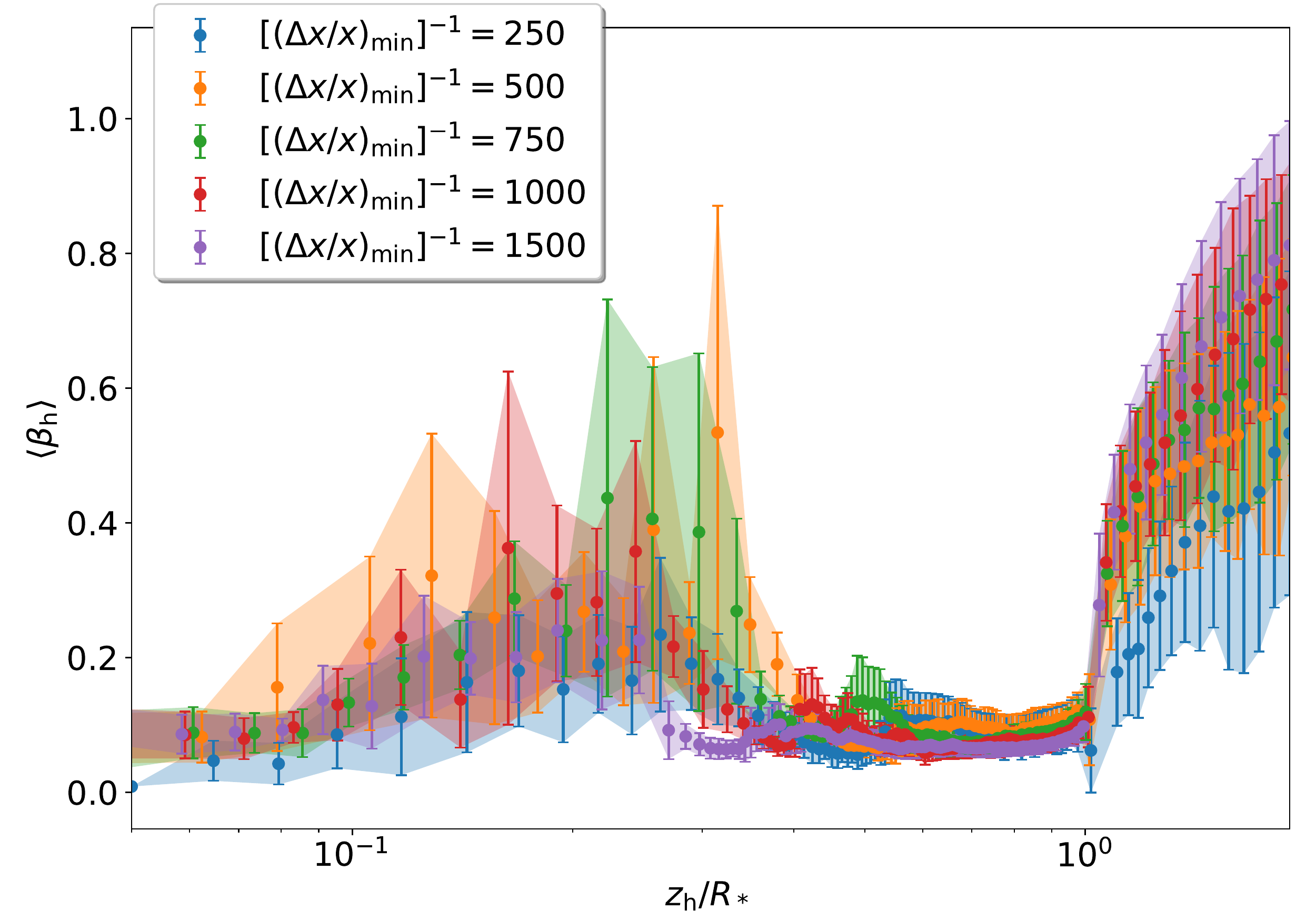}
    \caption{Results of the resolution test showing the average head forward shock velocity as a function of the head position $z$ for the simulation with $\theta_\jet = 0.1~\rad$ and $L_\jet = 10^{51}\erg~\s^{-1}$ at different resolutions. 
    The resolutions are: $620 \times 595 $  (blue), $1240 \times 1190$ (orange), $1860 \times 1785$ (green), $2480 \times 2380$ (red),and D) $3720 \times 3570$ (purple). Convergence is excellent for $z_{\rm}/R_*> 0.5$. As the numerical initial conditions of the jets depend on the resolution, convergence is less clear for  $z_{\rm}/R_*< 0.5$ as in this regime the jet has to settle down and ``forget" its initial conditions (see text).}
    \label{fig: resolution_beta}
\end{figure}

\begin{figure}
    \centering
    \includegraphics[scale=0.34]{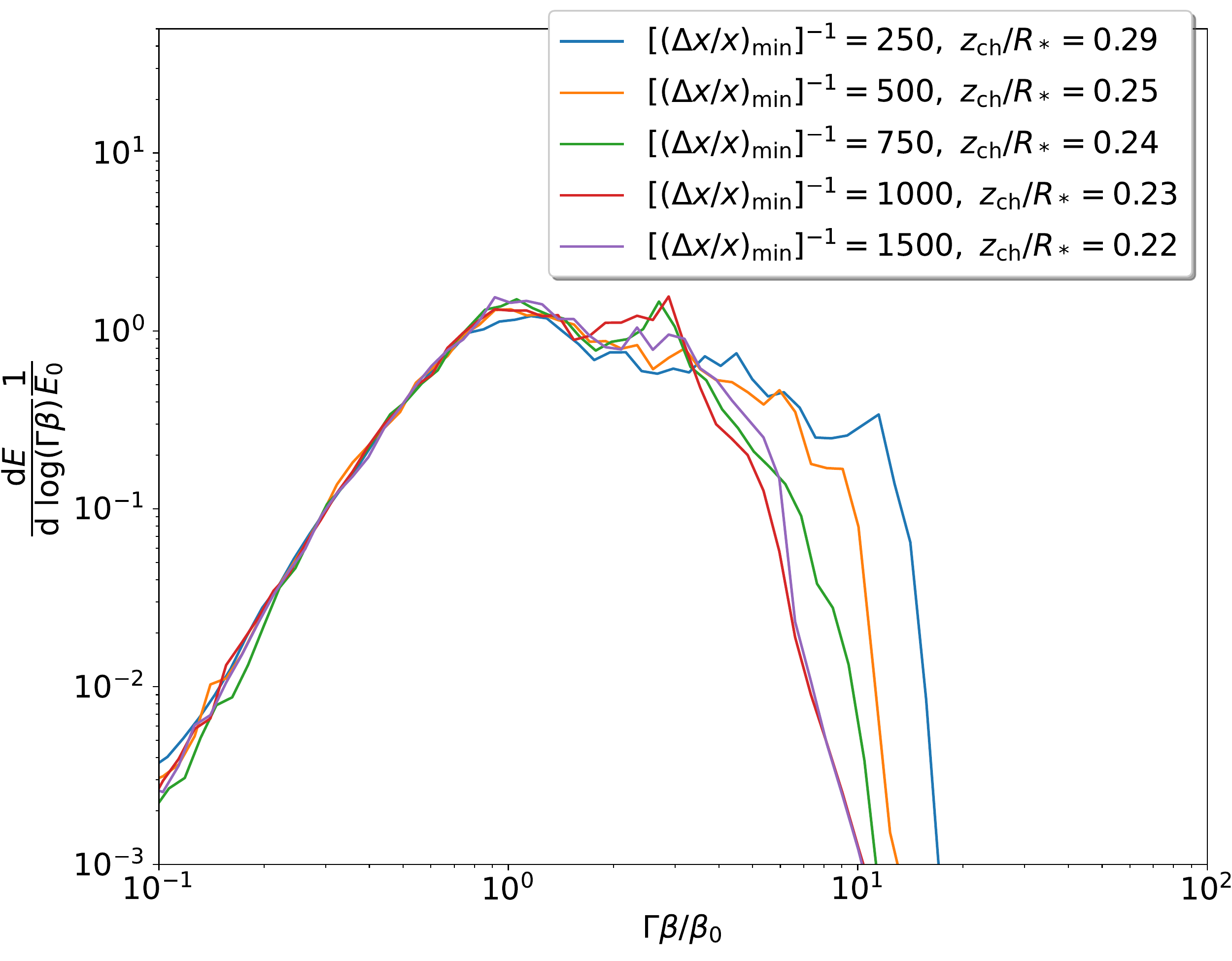}
    \caption{The energy-velocity distributions for  simulations with $\theta_\jet = 0.1~\rad$ and $L_\jet = 10^{51}\erg~\s^{-1}$ with different resolutions. For each curve we report the different choking height relative to the stellar radius. The resolutions are the same as reported for the curves in Fig.~\ref{fig: resolution_beta}.}
    \label{fig: resolution_energy_velocity}
\end{figure}

We tested the convergence of our jet evolution in 2D varying the grid resolution of our simulation box for the case $\theta_\jet = 0.1~\rad$ and $L_\jet = 10^{51}~\erg / \s$. 
The engine powering the jet in each simulation stops after 1 second and thus we expect that the jet is choked inside the stellar atmosphere at around the same height and the cocoon expands slowly before accelerating again close to the stellar edge. 

We tested our setup for different resolutions: 1) $620 \times 595 $, 2) $1240 \times 1190$, 3) $1860 \times 1785$, 4) $2480 \times 2380$, and 5) $3720 \times 3570$ (purple). 
Resolution 4) is what we use for all the simulations in this paper. 
For these different resolutions we investigated the head velocity convergence. 
Figure \ref{fig: resolution_beta} presents the head velocity as a function of time for the five resolutions previously discussed. 

We see how the lower resolution runs tend to propagate slightly slower than the high resolution runs (red and purple dots) before the jet engine terminates its activity. 
Low resolution runs also tend to converge slower to a constant velocity in the internal part of the the stellar atmosphere after the jet engine switches off and the choking happens. 
This results in a different choking height position $z_\mathrm{ch}$ for the low resolution runs with respect to the high resolution cases (red and purple). 
At around $R \simeq R_*/2$ the head velocity stabilizes and reaches a stable value of $0.06~\mathrm{c}$ for all the resolutions taken into account. The convergence to this velocity is excellent, however it takes some time for the jet to reach this velocity and ``forget" about its initial conditions, which depends on the resolution.
After the forward shock reaches $R_\ast/2$ the head accelerates as it nears the edge of the star, eventually resulting in a shock breakout at $z_\mathrm{ch}/R_* = 1$. 
After the shock breakout we notice that the forward shock propagation for the low resolution runs is much slower with respect to the high resolution simulations, a clear sign that the numerical resolution is insufficient to treat the problem properly. 

In Fig.~\ref{fig: resolution_energy_velocity} we plot the energy-velocity distribution for the five different resolutions we analyzed for Fig.~\ref{fig: resolution_beta}. We can clearly see how the distribution converges to a similar shape (red and purple curves at the highest resolutions) at the resolution increases. The two highest resolutions differ very little qualitatively and the jets are choked almost at the same $z_\choke$. This demonstrates that our results are sufficiently converged for the runs used in the paper.

\section{Different smoothing function}
\label{sec: appendix B}

\begin{figure}
    \centering
    \includegraphics[scale=0.33]{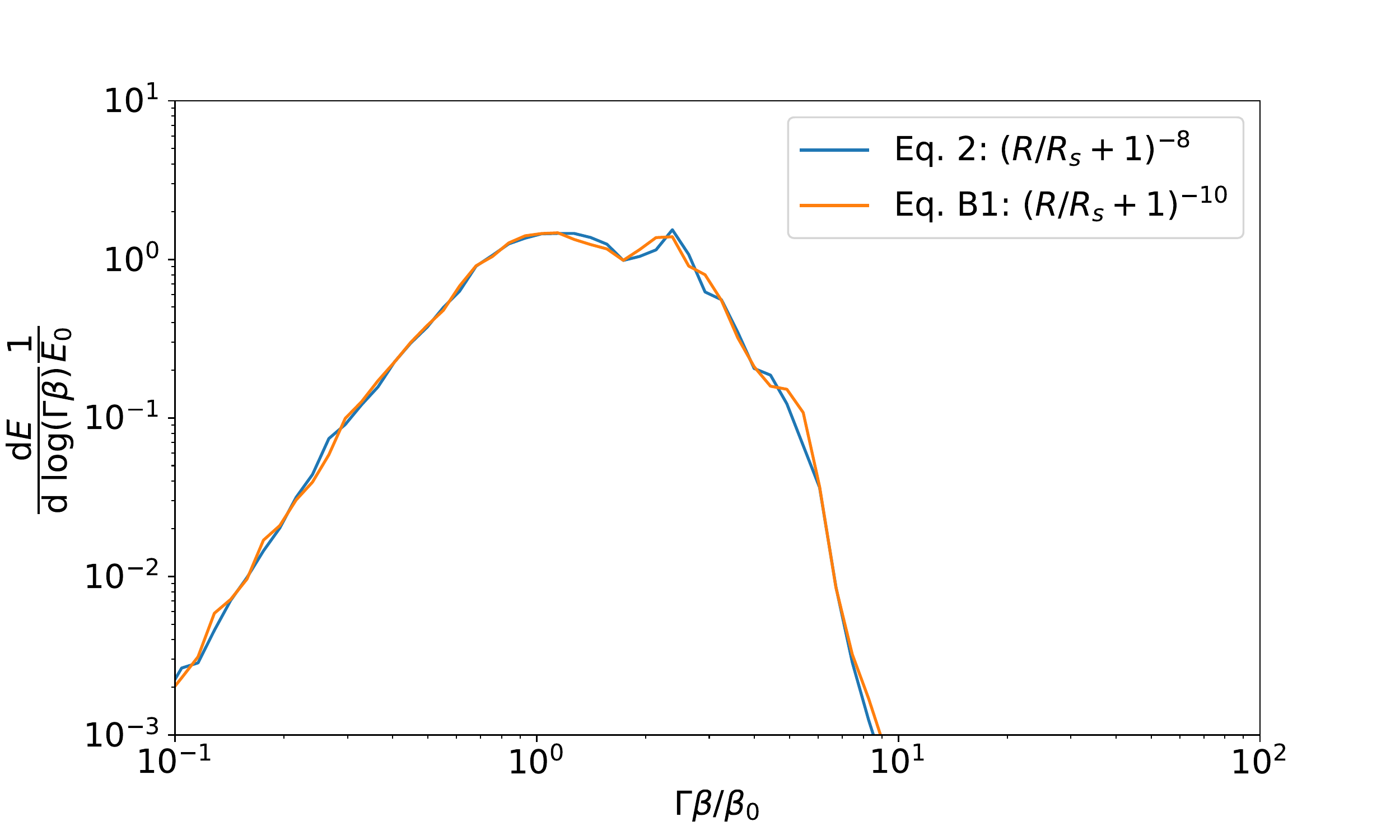}
    \caption{Comparison of the energy-velocity distribution of two simulations of a jet with canonical parameters with two different smoothing functions. 
    The blue line uses the smoothing function of Eq.~\ref{eq: smooth}, while the orange line uses the smoothing of Eq.~\ref{eq: smooth2}.}
    \label{fig: different_smoothing}
\end{figure}

The smoothing function for the stellar density profile (Eq.~\ref{eq: smooth}) prevents the computation to break down at the moment of the shock breakout because of the sudden drop in density of $\sim$20 orders of magnitude at the edge of the star in a relatively small distance. 
The functional form of the smoothing function  however is either arbitrary, so we tested whether the result is affected by the particular choice of  Eq.~\ref{eq: smooth}. 
We run a simulation for canonical jet parameters with the sharper smoothing function
\begin{equation}
\label{eq: smooth2}
    \rho_\mathrm{smooth, 2} (R) = \rho_\mathrm{s} \left( \dfrac{R}{R_\mathrm{s}} + 1\right)^{-10} \ , 
\end{equation}
with $\rho_\mathrm{s} = 0.05~\g~\cm^{-3}$ and $R_\mathrm{s} = 5 \times 10^{8}~\cm$.
In Fig.~\ref{fig: different_smoothing} we report the comparison of the energy-velocity distribution of this new run with the old one which uses the smoothing of Eq.~\ref{eq: smooth}. 
We can immediately see that the the two curves almost overlap and show insignificant differences which do not affect the final result. 
This is also due to the fact that the external additional mass given by these smoothing functions is of the order of $10^{-6}~M_*$, which is completely negligible and do not alter the physical scale of the system.


\bsp	
\label{lastpage}
\end{document}